\pdfoutput=1 

\newif\ifdraft\draftfalse
\newif\ifsoon\soontrue
\newif\iffull\fullfalse
\newif\ifanon\anonfalse
\newif\ifdesperateforspace\desperateforspacefalse
\newif\iflater\laterfalse
\newif\ifreadable\readablefalse
\newif\ifpostsubmission\postsubmissiontrue 
\makeatletter \@input{texdirectives} \makeatother

\ifpostsubmission{}
\documentclass{llncs}
\else{}
\documentclass[compsoc, conference, letterpaper, 10pt]{IEEEtran}
\fi{}

\usepackage{times}

\usepackage{hyperref} 
\hypersetup{breaklinks} 
\usepackage[utf8]{inputenc}
\usepackage{listings}
\usepackage{mathpartir}
\usepackage[american]{babel}
\usepackage{amssymb}
\usepackage{amsmath}
\usepackage{inconsolata}
\usepackage{fullminipage}

\usepackage{cite}

\ifpostsubmission
\else
\usepackage{amsthm}

\theoremstyle{plain}
\newtheorem{theorem}{Theorem}[section]

\newtheorem{corollary}[theorem]{Corollary}

\theoremstyle{definition}

\theoremstyle{remark}

\fi

\usepackage{array}
\usepackage{xspace}
\usepackage{cleveref}
\crefname{principle}{principle}{principles}
\crefformat{section}{\S#2#1#3}
\crefmultiformat{section}{\S\S#2#1#3}{and~#2#1#3}{, #2#1#3}{, and~#2#1#3}

\usepackage{stmaryrd}
\usepackage{xcolor}
\usepackage[shortlabels,inline]{enumitem}
\usepackage[autostyle]{csquotes}
\usepackage[normalem]{ulem}
\usepackage{combelow}

\usepackage[\ifdraft draft,\fi margin=false,inline=true]{fixme}
\definecolor{dkpurple}{rgb}{0.7,0,0.4}
\definecolor{dkgreen}{rgb}{0,0.4,0}
\makeatletter
\renewcommand*\FXLayoutInline[3]{%
  {\@fxuseface{inline}\ignorespaces #3 \fxnotename{#1}: #2]}}
\makeatother
\FXRegisterAuthor{bcp}{anbcp}{\color{dkpurple}[BCP}
\FXRegisterAuthor{aaa}{anaaa}{\color{teal}[AAA}
\FXRegisterAuthor{apt}{anapt}{\color{dkgreen}[APT} 
\FXRegisterAuthor{ch}{anch}{\color{dkgreen}[CH}
\FXRegisterAuthor{asz}{anasz}{\color{cyan}[ASZ}
\FXRegisterAuthor{yj}{anyj}{\color{violet}[YJ} 

\def\bcp{\bcpnote}
\def\aaa{\aaanote}

\def\ch{\chnote}
\let\asz\asznote

\newcommand*{\EG}{e.g.,\xspace}
\newcommand*{\IE}{i.e.,\xspace}
\newcommand*{\ETAL}{{\em et al.}\xspace}

\nonstopmode

\newcommand{\triple}[3]{\{#1\}\; #2\; \{#3\}}

\newcommand{\codeface}[1]{\mathsf{#1}}
\newcommand{\ctrue}{\codeface{true}}
\newcommand{\cfalse}{\codeface{false}}

\newcommand{\freshf}{\codeface{fresh}}

\newcommand{\cskip}{\codeface{skip}}
\newcommand{\calloc}{\codeface{alloc}}
\newcommand{\cfree}{\codeface{free}}

\newcommand{\cif}{\codeface{if}}
\newcommand{\cthen}{\codeface{then}}
\newcommand{\celse}{\codeface{else}}
\newcommand{\cwhile}{\codeface{while}}
\newcommand{\cdo}{\codeface{do}}
\newcommand{\cend}{\codeface{end}}
\newcommand{\nil}{\codeface{nil}}
\newcommand{\oerror}{\codeface{error}}

\newcommand{\cast}{\codeface{cast}}

\newcommand{\cifte}[3]{\cif\; #1 \;\cthen\; #2 \;\celse\; #3}
\newcommand{\cwhiledo}[2]{\cwhile\; #1 \;\cdo\; #2 \;\cend}
\newcommand{\Nop}{\mathsf{Nop}}
\newcommand{\Const}{\mathsf{Const}}
\newcommand{\Mov}{\mathsf{Mov}}
\newcommand{\Binop}{\mathsf{Binop}}
\newcommand{\Load}{\mathsf{Load}}
\newcommand{\Store}{\mathsf{Store}}
\newcommand{\Jump}{\mathsf{Jump}}
\newcommand{\Jal}{\mathsf{Jal}}
\newcommand{\Bnz}{\mathsf{Bnz}}
\newcommand{\Halt}{\mathsf{Halt}}

\newcommand{\cand}{\mathbin{\codeface{and}}}
\newcommand{\cor}{\mathbin{\codeface{or}}}
\newcommand{\cnot}{\codeface{not}}
\newcommand{\coffset}{\codeface{offset}}

\newcommand{\var}{\mathsf{var}}

\newcommand{\Z}{\mathbb{Z}}

\newcommand{\I}{\mathbb{I}}
\newcommand{\B}{\mathbb{B}}

\newcommand{\St}{\mathcal{S}}
\newcommand{\Ls}{\mathcal{L}}
\newcommand{\M}{\mathcal{M}}
\newcommand{\V}{\mathcal{V}}
\newcommand{\Ot}{\mathcal{O}}
\newcommand{\W}{\mathcal{W}}
\newcommand{\R}{\mathcal{R}}

\newcommand{\lsb}{\llbracket}
\newcommand{\rsb}{\rrbracket}
\newcommand{\dom}{\mathsf{dom}}

\newcommand{\bind}{\mathsf{bind}}

\newcommand{\fin}{\mathrm{fin}}

\newcommand{\modvars}{\mathsf{modvars}}
\newcommand{\independent}{\mathsf{independent}}
\newcommand{\partfun}{\rightharpoonup}
\newcommand{\partfunfin}{\partfun_\fin}
\newcommand{\power}{\mathcal{P}}
\newcommand{\vars}{\mathsf{vars}}
\newcommand{\blocks}{\mathsf{blocks}}
\newcommand{\ids}{\mathsf{ids}}
\newcommand{\finalids}{\mathsf{finalids}}

\newcommand{\fresh}{\mathrel{\#}}
\newcommand{\teq}{\triangleq}

\begin{document}

\title{\huge The Meaning of Memory Safety\ifanon\vspace{-1em}\fi}

\ifpostsubmission
\author{
\ifanon
\else
  Arthur Azevedo de Amorim\inst{1} \and 
  C\u{a}t\u{a}lin Hri\cb{t}cu\inst{2} \and 
  Benjamin C. Pierce\inst{3} 
\fi
}
\institute{
\ifanon
\else
  Carnegie Mellon University \and
  Inria Paris\and
  University of Pennsylvania
\fi{}
}

\fi

\maketitle

\aaa{Make sure that the ``draft'' option of hyperref can be removed}

\begin{abstract}
We give a rigorous characterization of what it means for a programming language
to be {\em memory safe}, capturing the intuition that memory safety
supports {\em local reasoning about state}.
We formalize this principle in two ways.  First, we show how a small memory-safe
language validates a {\em noninterference} property: a program can neither
affect nor be affected by unreachable parts of the state.  Second,
we extend separation logic, a proof system for
heap-manipulating programs, with a ``memory-safe variant'' of its {\em frame
  rule}.  The new
rule is stronger because it applies even when parts of the program are buggy or
malicious, but also weaker because it demands a stricter form of separation
between parts of the program state.  We also consider a number of pragmatically
motivated variations on memory safety and the reasoning principles they support.
As an application of our characterization, we evaluate the security of a
previously proposed dynamic monitor for memory safety of heap-allocated data.
\end{abstract}

%

\section{Introduction}
\label{sec:introduction}

Memory safety, and the vulnerabilities that follow from its absence~\cite{Szekeres2013}, are common concerns.
So what {is} it, exactly?  Intuitions abound, but translating
them into satisfying formal definitions is surprisingly
difficult~\cite{Hicks:memory-safety}.

In large part, this difficulty stems from the prominent role that informal,
everyday intuition assigns, in discussions of memory safety, to a range of
errors related to memory {\em mis}use---buffer overruns, double frees, etc.
Characterizing memory safety in terms of the absence of these errors is
tempting, but this falls short for two reasons.  First, there is often
disagreement on which behaviors qualify as errors.  For example, many real-world
C programs intentionally rely on unrestricted pointer
arithmetic~\cite{MemarianMLNCWS16}, though it may yield undefined behavior
according to the language standard~\cite[§6.5.6]{ISO:C99}. Second, from the
perspective of security, the critical issue is not the errors themselves, but
rather the fact that, when they occur in unsafe languages like C, the
program's ensuing behavior is determined by obscure, low-level factors such as
the compiler's choice of run-time memory layout, often leading to exploitable
vulnerabilities.  By contrast, in memory-safe languages like Java, programs
can attempt to access arrays out of bounds, but such mistakes lead to
sensible, predictable outcomes.

Rather than attempting a definition in terms of bad things that cannot happen,
we aim to formalize memory safety in terms of {\em reasoning principles} that
programmers can soundly apply in its presence (or conversely, principles that
programmers should {\em not} naively apply in unsafe settings, because doing so
can lead to serious bugs and vulnerabilities).  Specifically, to give an account
of {\em memory} safety, as opposed to more inclusive terms such as ``type
safety,''\ch{Could cite this paper
  \url{http://lucacardelli.name/Papers/TypeSystems.pdf} for a careful discussion
  of what type safety means in terms of preventing errors} we focus on reasoning
principles that are common to a wide range of stateful abstractions, such as
records, tagged or untagged unions, local variables, closures, arrays, call
stacks, objects, compartments, and address spaces.

What sort of reasoning principles?  Our inspiration comes from {\em
  separation logic}~\cite{Reynolds:2002}, a variant of Hoare logic designed to
verify complex heap-manipulating programs.  The power of separation logic stems
from \emph{local reasoning} about state: to prove the correctness of a program
component, we must argue that its memory accesses are confined to a
\emph{footprint}, a precise region demarcated by the specification.  This
discipline allows proofs to ignore regions outside of the footprint, while
ensuring that {arbitrary} invariants for these regions are preserved
during execution.

The locality of separation logic is deeply linked to memory safety.  Consider a
hypothetical jpeg decoding procedure that manipulates image buffers.  We might
expect its execution not to interfere with the integrity of an unrelated window
object in the program.  We can formalize this requirement in separation logic by
proving a specification that includes only the image buffers, but not the
window, in the decoder's footprint. Showing that the footprint is respected
would amount to checking the bounds of individual buffer accesses, thus
enforcing memory safety; conversely, if the decoder is not memory safe, a simple
buffer overflow might suffice to tamper with the window object, thus violating
locality and potentially paving the way to an attack.

Our aim is to extend this line of reasoning beyond conventional separation
logic, seeking to encompass settings such as ML, Java, or Lisp that enforce memory
safety automatically without requiring complete correctness proofs---which can
be prohibitively expensive for large code bases, especially in the presence of
third-party libraries or plugins over which we have little control.  The key
observation is that memory safety forces code to respect a natural footprint:
the set of its reachable memory locations (reachable with respect to the
variables it mentions).  Suppose that the jpeg decoder above is written in Java.
Though we may not know much about its input-output behavior, we can still assert
that it cannot have any effect on the window object simply by replacing the
detailed reasoning demanded by separation logic by a simple inaccessibility
check.

Our \emph{first contribution} is to formalize local reasoning principles
supported by an ideal notion of memory safety, using a simple language
(\Cref{sec:imp}) to ground our discussion.  We show three results
(\Cref{thm:frame-ok,thm:frame-loop,thm:frame-error}) that explain how the
execution of a piece of code is affected by extending its initial heap.  These
results lead to a \emph{noninterference} property (\Cref{cor:noninterference}),
ensuring that code cannot affect or be affected by unreachable memory.  In
\Cref{sec:separation}, we show how these results yield a variant of the frame
rule of separation logic (\Cref{thm:weak-frame-rule}), which embodies its local
reasoning capabilities. The two variants have complementary strengths and
weaknesses: while the original rule applies to unsafe settings like C, but
requires comprehensively verifying individual memory accesses, our variant does
not require proving that every access is correct, but demands a stronger notion
of separation between memory regions.
These results have been verified with the Coq proof
assistant.\footnote{The proofs are available \ifanon
  at \url{https://github.com/CCS492/MemorySafety}.  \else at:
  \url{https://github.com/arthuraa/memory-safe-language}.\fi}

Our \emph{second contribution} (\Cref{sec:relaxations}) is to evaluate
pragmatically motivated relaxations of the ideal notion above, exploring various
trade-offs between safety, performance, flexibility, and backwards
compatibility.  These variants can be broadly classified into two groups
according to reasoning principles they support.  The stronger group gives up on
some secrecy guarantees, but still ensures that pieces of code cannot modify the
contents of unreachable parts of the heap.  The weaker group, on the other hand,
leaves gaps that completely invalidate reachability-based reasoning.

Our \emph{third contribution} (\Cref{sec:micro-policy}) is to demonstrate how
our characterization applies to more realistic settings, by analyzing a
heap-safety monitor for machine code~\cite{pump_asplos2015,micropolicies2015}.
We prove that the abstract machine that it implements also satisfies a
noninterference property, which can be transferred to the monitor via
refinement, modulo memory exhaustion issues discussed in \Cref{sec:relaxations}.
These proofs are also done in Coq.\footnote{
Available at \url{https://github.com/micro-policies/micro-policies-coq/tree/master/memory_safety}.}

\ifpostsubmission{}\fi{}

We discuss related work on memory safety and stronger reasoning principles in
\Cref{sec:related-work}, and conclude in \Cref{sec:conclusion}.
While memory safety has seen prior formal investigation
(e.g. \cite{NagarakatteZMZ09,SwamyHMGJ06}),
%
%
our characterization is the first phrased in terms of reasoning principles that
are valid when memory safety is enforced automatically.
We hope that these principles can serve as good criteria for formally evaluating
such enforcement mechanisms in practice.
Moreover, our definition is self-contained and does not rely on additional
features such as full-blown capabilities, objects, module systems, etc.  Since
these features tend to depend on some form of memory safety anyway, we could see
our characterization as a common core of reasoning principles that underpin all
of them.

\section{An Idealized Memory-Safe Language}
\label{sec:imp}

Our discussion begins with a concrete case study: a simple imperative language
with manual memory management.  It features several mechanisms for controlling
the effects of memory misuse, ranging from the most conventional, such as bounds
checking for spatial safety, to more uncommon ones, such as assigning unique
identifiers to every allocated block for ensuring temporal safety.

Choosing a language with manual memory management may seem odd, since safety is
often associated with garbage collection.  We made this choice for two reasons.
First, most discussions on memory safety are motivated by its absence from
languages like C that also rely on manual memory management.  There is a vast
body of research that tries to make such languages safer, and we would like our
account to apply to it.  Second, we wanted to stress that our characterization
does not depend fundamentally on the mechanisms used to enforce memory safety,
especially because they might have complementary advantages and shortcomings.
For example, manual memory management can lead to more memory leaks; garbage
collectors can degrade performance; and specialized type systems for managing
memory~\cite{SwamyHMGJ06,Rust} are more complex. After a brief overview of the
language, we explore its reasoning principles in \Cref{sec:reasoning}.

\newcommand{\statetableshort}{
  \begin{alignat*}{2}
  s \in \St & \teq \Ls \times \M && \text{ (states)} \\
  l \in \Ls & \teq \var \partfunfin \V && \text{ (local stores)} \\
  m \in \M  & \teq \I \times \Z \partfunfin \V && \text{ (heaps)} \\
  v \in \V  & \teq \Z \uplus \B \uplus \{\nil\} \uplus \I \times \Z && \text{ (values)} \\
  \Ot & \teq \St \uplus \{ \oerror \} && \text{ (outcomes)}
\end{alignat*}
\begin{align*}
  \I  & \teq \text{a countably infinite set} \\
  X \partfunfin Y & \teq \text{finite partial functions $X \partfun Y$}
\end{align*}
}

\newcommand{\statetablelong}{
\begin{align*}
  s \in \St & \teq \Ls \times \M & \text{(states)} \\
  l \in \Ls & \teq \var \partfunfin \V & \text{(local stores)} \\
  m \in \M  & \teq \I \times \Z \partfunfin \V & \text{(heaps)} \\
  v \in \V  & \teq \Z \uplus \B \uplus \{\nil\} \uplus \I \times \Z & \text{(values)} \\
  \Ot & \teq \St \uplus \{ \oerror \} & \text{(outcomes)}
\end{align*}
\begin{align*}
  \I  & \teq \text{some countably infinite set} \\
  X \partfunfin Y & \teq \text{partial functions $X \partfun Y$ with finite domain}
\end{align*}}

\begin{figure}[t]
\centering

\begin{minipage}{0.45\linewidth}
\begin{tabular}{r|l}
  Command                     & Description         \\ \hline
  $x \gets e$                 & Local assignment \\
  $x \gets [e]$               & Read from heap      \\
  $[e_1] \gets e_2$           & Heap assignment     \\
  $x \gets \calloc(e_{size})$ & Allocation          \\
  $\cfree(e)$                 & Deallocation        \\
  $\cskip$                    & Do nothing          \\
  $\cifte{e}{c_1}{c_2}$       & Conditional         \\
  $\cwhiledo{e}{c}$           & Loop                \\
  $c_1; c_2$                  & Sequencing
\end{tabular}
\end{minipage}
\begin{minipage}{0.45\linewidth}
\statetableshort
\end{minipage}

\caption{Syntax, states and values}
\label{fig:syntax-and-states}
\end{figure}

\Cref{fig:syntax-and-states} summarizes the language syntax and other basic
definitions.  Expressions $e$ include variables $x \in \var$, numbers
$n \in \Z$, booleans $b \in \B$, an invalid pointer $\nil$, and various
operations, both binary (arithmetic, logic, etc.)  and unary (extracting the
offset of a pointer). We write $[e]$ for dereferencing the pointer denoted by
$e$.

Programs operate on states consisting of two components: a \emph{local store},
which maps variables to values, and a \emph{heap}, which maps pointers to
values.  Pointers are not bare integers, but rather pairs $(i, n)$ of a
\emph{block identifier} $i \in \I$ and an offset $n \in \Z$.  The offset is
relative to the corresponding block, and the identifier $i$ need not bear any
direct relation to the physical address that might be used in a concrete
implementation on a conventional machine.  (That is, we can equivalently think
of the heap as mapping each identifier to a separate array of heap cells.)
Similar structured memory models are widely used in the literature, as in the
CompCert verified C compiler~\cite{LeroyB08} and other models of the C
language~\cite{KangHMGZV15}, for instance.

We write $\lsb c \rsb(s)$ to denote the outcome of running a program $c$ in an
initial state $s$, which can be either a successful final state $s'$ or a fatal
run-time error.  Note that $\lsb c \rsb$ is partial, to account for
non-termination.  Similarly, $\lsb e \rsb (s)$ denotes the result of evaluating
the expression $e$ on the state $s$ (expression evaluation is total and has no
side effects).  The formal definition of these functions is left to the
Appendix\iflater\bcp{Recommend integrating and explaining it here---the
  technicalities that follow are hard to understand without a clear picture of
  how the machine works.  Probably the best way is to put it in a subsection at
  the end of this section}\ch{The current section does read well on its own, and
  more details here wouldn't have helped me in understanding 3 better, so I
  think leaving the details for the appendix is good. In general, not missing
  details are the main problem for understanding 3, but missing intuition.}\fi;
we just single out a few aspects that have a crucial effect on the security
properties discussed later.

\paragraph*{Illegal Memory Accesses Lead to Errors}

The language controls the effect of memory misuse by raising errors that stop
execution immediately.  This contrasts with typical C implementations, where
such errors lead to unpredictable \emph{undefined behavior}.
The main errors are caused by reads, writes, and frees to the current memory $m$
using \emph{invalid pointers}---that is, pointers $p$ such that $m(p)$ is
undefined.  Such pointers typically arise by offsetting an existing pointer out
of bounds or by freeing a structure on the heap (which turns all other pointers
to that block in the program state into dangling ones).  In common parlance,
this discipline ensures both \emph{spatial} and \emph{temporal} memory safety.

\paragraph*{Block Identifiers are Capabilities}

Pointers can only be used to access memory corresponding to their identifiers,
which effectively act as capabilities. Identifiers are set at allocation time,
where they are chosen to be fresh with respect to the entire current state (\IE
the new identifier is not associated with any pointers defined in the current
memory, stored in local variables, or stored on the heap).  Once assigned,
identifiers are immutable, making it impossible to fabricate a pointer to an
allocated block out of thin air.  This can be seen, for instance, in the
semantics of addition, which allows pointer arithmetic but does not affect
identifiers: %
\ifpostsubmission{}
\begin{align*}
  \lsb e_1 + e_2 \rsb(s) & \teq
  \begin{cases}
    n_1 + n_2 & \text{if $\lsb e_i\rsb(s) = n_i$} \\
    (i,n_1+n_2) & \text{if $\lsb e_1\rsb(s) = (i,n_1)$ and $\lsb e_2\rsb(s) = n_2$} \\
    \nil & \text{otherwise}
  \end{cases}
\end{align*}
\fi{}%
For simplicity, nonsensical combinations such as adding two pointers simply
result in the $\nil$ value. A real implementation might represent identifiers
with hardware tags and use an increasing counter to generate identifiers for new
blocks (as done by Dhawan \ETAL \cite{pump_asplos2015}; see \cref{sec:monitor});
if enough tags are available, every identifier will be fresh.

\paragraph*{Block Identifiers Cannot be Observed}
Because of the freshness condition above, identifiers can reveal information
about the entire program state.  For example, if they are chosen according to an
increasing counter, knowing what identifier was assigned to a new block tells us
how many allocations have been performed.  A concrete implementation would face
similar issues related to the choice of physical addresses for new
allocations. (Such issues are commonplace in systems that combine dynamic
allocation and information-flow control~\cite{AmorimCDDHPPPT16}.)
%
For this reason, our language keeps identifiers opaque and inaccessible to
programs; they can only be used to reference values in memory, and nothing else.
We discuss a more permissive approach\ifpostsubmission{}\fi{} in \Cref{sec:observing-pointers}.

Note that hiding identifiers doesn't mean we have to hide \emph{everything}
associated with a pointer: besides using pointers to access memory, programs can
also safely extract their offsets and test if two pointers are equal (which
means equality for both offsets and identifiers).  Our Coq development also
shows that it is sound to compute the size of a memory block via a valid
pointer.

\paragraph*{New Memory is Always Initialized}


Whenever a memory block is allocated, all of its contents are initialized to
$0$.  (The exact value does not matter, as long it is some constant that is not
a previously allocated pointer.)  This is important for ensuring that allocation
does not leak secrets present in previously freed blocks; we return to this
point in \Cref{sec:uninitialized}.

\section{Reasoning with Memory Safety}
\label{sec:reasoning}

Having presented our language, we now turn to the reasoning principles that it
supports.  Intuitively, these principles allow us to analyze the effect of a
piece of code by restricting our attention to a smaller portion of the program
state.  A first set of \emph{frame theorems} (\ref{thm:frame-ok},
\ref{thm:frame-loop}, and \ref{thm:frame-error}) describes how the execution of
a piece of code is affected by extending the initial state on which it runs.
These in turn imply a noninterference property, \Cref{cor:noninterference},
guaranteeing that program execution is independent of inaccessible memory
regions---that is, those that correspond to block identifiers that a piece of
code does not possess.  Finally, in \Cref{sec:separation}, we discuss how the
frame theorems can be recast in the language of separation logic, leading to a
new variant of its frame rule (\Cref{thm:weak-frame-rule}).

\subsection{Basic Properties of Memory Safety}
\label{sec:frame-theorems}

\begin{figure}[t]
\begin{alignat*}{2}
  (l_1,m_1) \cup (l_2, m_2) & \teq (l_1 \cup l_2, m_1 \cup m_2) && \text{ (state union)}\\
  (f \cup g)(x) & \teq
  \begin{cases}
    f(x) & \text{if $x \in \dom(f)$} \\
    g(x) & \text{otherwise}
  \end{cases} && \text{ (partial function union)} \\
  \blocks(l, m) & \teq \{ i \in \I \mid \exists n, (i, n) \in \dom(m) \}
  && \text{ (identifiers of live blocks)} \\
  \ids(l, m) & \teq \blocks(l, m) && \text{ (all identifiers in state)} \\
  & \cup\{ i \mid \exists x, n, l(x) = (i, n) \} && \\
  & \cup \{ i \mid \exists p, n, m(p)  = (i, n) \} && \\
  \vars(l, m) & \teq \dom(l) && \text{ (defined local variables)} \\
  \vars(c) & \teq \text{local variables of program $c$} && \\
  X \fresh Y & \teq (X \cap Y = \emptyset) && \text{ (disjoint sets)} \\
  \pi \cdot s & \teq \text{rename identifiers with permutation $\pi$}
\end{alignat*}
\caption{Basic notation}
\label{fig:basic-notation}
\end{figure}

\Cref{fig:basic-notation} summarizes basic notation used in our results.  By
\emph{permutation}, we mean a function $\pi : \I \to \I$ that has a two-sided
inverse $\pi^{-1}$; that is,
$\pi \circ \pi^{-1} = \pi^{-1} \circ \pi = \mathsf{id}_{\I}$. Some of these
operations are standard and omitted for brevity.\footnote{%
  The renaming operation $\pi \cdot s$, in particular, can be derived formally
  by viewing $\St$ as a nominal set over $\I$~\cite{Pitts:2013} obtained by
  combining products, disjoint unions, and partial functions.}

The first frame theorem states that, if a program terminates successfully, then
we can extend its initial state almost without affecting execution.
\begin{theorem}[Frame OK]
  \label{thm:frame-ok}
  Let $c$ be a command, and $s_1$, $s_1'$, and $s_2$ be states. Suppose that
  $\lsb c\rsb(s_1) = s_1'$, $\vars(c) \subseteq \vars(s_1)$, and
  $\blocks(s_1) \fresh \blocks(s_2)$.  Then there exists a permutation $\pi$
  such that $\lsb c\rsb(s_1 \cup s_2) = \pi \cdot s_1' \cup s_2$ and
  $\blocks(\pi \cdot s_1') \fresh \blocks(s_2)$.
\end{theorem}
The second premise, $\vars(c) \subseteq \vars(s_1)$, guarantees that all the
variables needed to run $c$ are already defined in $s_1$, implying that their
values do not change once we extend that initial state with $s_2$.  The third
premise, $\blocks(s_1) \fresh \blocks(s_2)$, means that the memories of $s_1$
and $s_2$ store disjoint regions.  Finally, the conclusion of the theorem states
that (1) the execution of $c$ does not affect the extra state $s_2$ and (2) the
rest of the result is almost the same as $s_1'$, except for a permutation of
block identifiers.

Permutations are needed to avoid clashes between identifiers in $s_2$ and those
assigned to regions allocated by $c$ when running on $s_1$.  For instance,
suppose that the execution of $c$ on $s_1$ allocated a new block, and that this
block was assigned some identifier $i \in \I$.  If the memory of $s_2$ already
had a block corresponding to $i$, $c$ would have to choose a different
identifier $i'$ for allocating that block when running on $s_1 \cup s_2$.  This
change requires replacing all occurrences of $i$ by $i'$ in the result of the
first execution, which can be achieved with a permutation that swaps these two
identifiers.%
\ifpostsubmission{}\fi{}

The proof of \Cref{thm:frame-ok} relies crucially on the facts that programs
cannot inspect identifiers, that memory can grow indefinitely (a common
assumption in formal models of memory), and that memory operations fail on
invalid pointers.  Because of the permutations, we also need to show that
permuting the initial state $s$ of a command $c$ with any permutation $\pi$
yields the same outcome, up to some additional permutation $\pi'$ that again
accounts for different choices of fresh identifiers.

\begin{theorem}[Renaming states]
  \label{thm:renaming}
  Let $s$ be a state, $c$ a command, and $\pi$ a permutation.  There exists
  $\pi'$ such that:
  \begin{mathpar}
  \lsb c\rsb(\pi \cdot s)
  =
  \begin{cases}
    \oerror & \text{if $\lsb c\rsb(s) = \oerror$} \\
    \bot & \text{if $\lsb c\rsb(s) = \bot$} \\
    \pi' \cdot \pi \cdot s' & \text{if $\lsb c\rsb(s) = s'$}
  \end{cases}
  \end{mathpar}
\end{theorem}

A similar line of reasoning yields a second frame theorem, which says that we
cannot make a program terminate just by extending its initial state.
\begin{theorem}[Frame Loop]
  \label{thm:frame-loop}~\\
  Let $c$ be a command, and $s_1$ and $s_2$ be states. If
  $\lsb c\rsb(s_1) = \bot$, $\vars(c) \subseteq \vars(s_1)$, and
  $\blocks(s_1) \fresh \blocks(s_2)$, then $\lsb c\rsb(s_1 \cup s_2) = \bot$.
\end{theorem}

The third frame theorem shows that extending the initial state also preserves
erroneous executions.  Its statement is similar to the previous ones, but with a
subtle twist. In general, by extending the state of a program with a block, we
might turn an erroneous execution into a successful one---if the error was
caused by accessing a pointer whose identifier matches that new block.  To avoid
this, we need a different premise ($\ids(s_1) \fresh \blocks(s_2)$) preventing
any pointers in the original state $s_1$ from referencing the new blocks
in~$s_2$---which is only useful because our language prevents programs from
forging pointers to existing regions. Since $\blocks(s) \subseteq \ids(s)$, this
premise is stronger than the analogous ones in the preceding results.

\begin{theorem}[Frame Error]
  \label{thm:frame-error}~\\
  Let $c$ be a command, and $s_1$ and $s_2$ be states. If
  $\lsb c\rsb(s_1) = \oerror$, $\vars(c)\subseteq \vars(s_1)$, and
  $\ids(s_1) \fresh \blocks(s_2)$, then $\lsb c\rsb(s_1 \cup s_2) = \oerror$.
\end{theorem}

\subsection{Memory Safety and Noninterference}

The consequences of memory safety analyzed so far are intimately tied to the
notion of \emph{noninterference}~\cite{GoguenM82}.  In its most widely
understood sense, noninterference is a \emph{secrecy} guarantee: varying secret
inputs has no effect on public outputs.  Sometimes, however, it is also used to
describe \emph{integrity} guarantees: low-integrity inputs have no effect on
high-integrity outputs.  In fact, both guarantees apply to unreachable memory in
our language, since they do not affect code execution; that is, execution (1)
cannot modify these inaccessible regions (preserving their integrity), and (2)
cannot learn anything meaningful about them, not even their presence (preserving
their secrecy).

\begin{corollary}[Noninterference]
  \label{cor:noninterference}
  Let $s_1$, $s_{21}$, and $s_{22}$ be states and $c$ be a command. Suppose that
  $\vars(c) \subseteq \vars(s_1)$, that $\ids(s_1) \fresh \blocks(s_{21})$ and
  that $\ids(s_1) \fresh \blocks(s_{22})$.  When running $c$ on the extended
  states $s_1 \cup s_{21}$ and $s_1 \cup s_{22}$, only one of the following
  three possibilities holds: %
  \ifpostsubmission{}%
  (1) both executions loop
  ($\lsb c\rsb(s_1 \cup s_{21}) = \lsb c\rsb(s_1 \cup s_{22}) = \bot$); %
  (2) both executions terminate with an error
  ($\lsb c\rsb(s_1 \cup s_{21}) = \lsb c\rsb(s_1 \cup s_{22}) = \oerror$); or %
  (3) both executions successfully terminate without interfering with the
  inaccessible portions $s_{21}$ and $s_{22}$ (formally, there exists a state
  $s_1'$ and permutations $\pi_1$ and $\pi_2$ such that
  $\lsb c\rsb(s_1 \cup s_{2i}) = \pi_i \cdot s_1' \cup s_{2i}$ and
  $\ids(\pi_i \cdot s_1') \fresh \blocks(s_{2i})$, for $i = 1, 2$).%
  \fi{}
\end{corollary}

\ifpostsubmission{}
\fi{}

Noninterference is often formulated using an \emph{indistinguishability
  relation} on states, which expresses that one state can be obtained from the
other by varying its secrets.  We could have equivalently phrased the above
result in a similar way.  Recall that the hypothesis
$\ids(s_1) \fresh \blocks(s_2)$ means that memory regions stored in $s_2$ are
unreachable via $s_1$.  Then, we could call two states ``indistinguishable'' if
the reachable portions are the same (except for a possible permutation).  In
\Cref{sec:relaxations}, the connection with noninterference will provide a good
benchmark for comparing different flavors of memory safety.

\subsection{Memory Safety and Separation Logic}
\label{sec:separation}

We now explore the relation between the principles identified above, especially
regarding integrity,
%
and the local reasoning facilities of separation logic.  Separation logic
targets specifications of the form $\triple{p}{c}{q}$, where $p$ and $q$ are
predicates over program states (subsets of $\St$). For our language, this could
roughly mean
\begin{align*}
  \forall s \in p, &\, \vars(c) \subseteq \vars(s) \Rightarrow \lsb c\rsb (s)
  \in q \cup \{\bot\}.
\end{align*}
That is, if we run $c$ in a state satisfying $p$, it will either diverge or
terminate in a state that satisfies $q$, but it will not trigger an error.  Part
of the motivation for precluding errors is that in unsafe settings like C they
yield undefined behavior, destroying all hope of verification.

Local reasoning in separation logic is embodied by the \emph{frame rule}, a
consequence of \Cref{thm:frame-ok,thm:frame-loop}.  Roughly, it says that a
verified program can only affect a well-defined portion of the state, with all
other memory regions left untouched.%
\footnote{Technically, the frame rule requires a slightly stronger notion of
  specification, accounting for permutations of allocated identifiers; our Coq
  development has a more precise statement.  \iflater\aaa{TODO: Include
    statement in appendix?}\bcp{OK to omit for now, I think.}\fi }

\begin{theorem}
\label{thm:frame-rule}
Let $p$, $q$, and $r$ be predicates over states and $c$ be a command.  The rule
\[
\inferrule*[Right=Frame]
  { \independent(r, \modvars(c)) \\ \triple{p}{c}{q} }
  { \triple{p * r}{c}{q * r} }
\]
is sound, where $\modvars(c)$ is the set of local variables modified by $c$,
$\independent(r, V)$ means that the assertion $r$ does not depend on the set of
local variables $V$
\[
\forall l_1\,l_2\,m, (\forall x \notin V, \; l_1(x) = l_2(x))
                     \Rightarrow (l_1, m) \in r \Rightarrow (l_2, m) \in r,
\]
and $p * r$ denotes the \emph{separating conjunction} of $p$ and $r$:
\[
  \{ (l, m_1 \cup m_2) \mid (l, m_1) \in p, (l, m_2) \in r,
                            \blocks(l, m_1) \fresh \blocks(l, m_2) \}.
\]
\end{theorem}

As useful as it is, precluding errors during execution makes it difficult to use
separation logic for \emph{partial verification}: proving {\em any} property, no
matter how simple, of a nontrivial program requires detailed reasoning about its
internals.  Even the following vacuous rule
is unsound in separation logic:
\[ \inferrule*[Right=Taut]{ }{\triple{p}{c}{\ctrue}} \]
\iflater\aaa{Should relate this to the safety theorem in the paper by
  Agten~\cite{Agten0P15}.}\fi%
For a counterexample, take $p$ to be $\ctrue$ and $c$ to be some arbitrary
memory read $x \gets [y]$.  If we run $c$ on an empty heap, which trivially
satisfies the precondition, we obtain an error, contradicting the specification.

Fortunately, our memory-safe language---in which errors\iflater
\ch{Only memory errors;
  a language can be memory safe while still having undefined behavior for things
  that are not memory related (e.g. integer overflows or whatever) ... although
  that's not the case for your simple language.  It seems that this whole
  sub-section talks about a special case, but that is not made explicit, which
  goes against the intro statement that ``we focus on reasoning principles that
  are directly related to mutable state''.  One local way to fix this would be
  to make the statement in this paragraph specific about your language, not
  about memory-safe languages in general [DONE], while maybe adding a footnote to
  discuss the more general case.}\fi{}
have a sensible, predictable semantics, as opposed to wild undefined
behavior---supports a variant of separation logic that allows looser
specifications of the form $\triple{p}{c}{q}_e$, defined as
\begin{align*}
  \forall s \in p, &\, \vars(c) \subseteq \vars(s) \Rightarrow \lsb c\rsb (s)
  \in q \cup \{\bot, \oerror\}.
\end{align*}

These specifications are weaker than their conventional counterparts, leading to
a subsumption rule:
\[ \inferrule{ \triple{p}{c}{q} }{ \triple{p}{c}{q}_e} \]

\ifpostsubmission{}%
Because errors are no longer prevented, the \textsc{Taut} rule
$\triple{p}{c}{\ctrue}_e$ \fi{}%
becomes sound, since the $\ctrue{}$ postcondition now means that any outcome
whatsoever is acceptable.  Unfortunately, there is a price to pay for allowing
errors: they compromise the soundness of the frame rule.  The reason, as hinted
in the introduction, is that preventing run-time errors has an additional
purpose in separation logic: it forces programs to act locally---that is, to
access only the memory delimited their pre- and postconditions.  To see why,
consider the same program $c$ as above, $x \gets [y]$.  This program clearly
yields an error when run on an empty heap, implying that the triple
\ifpostsubmission{}%
$\triple{\mathsf{emp}}{c}{x = 0}_e$ \fi{}%
is valid, where the predicate $\mathsf{emp}$ holds of any state with an empty
heap and $x = 0$ holds of states whose local store maps $x$ to $0$.  Now
consider what happens if we try to apply an analog of the frame rule to this
triple using the frame predicate $y \mapsto 1$, which holds in states where $y$
contains a pointer to the unique defined location on the heap, which stores the
value $1$.  After some simplification, we arrive at the specification
\ifpostsubmission{}%
$\triple{y \mapsto 1}{c}{x = 0 \wedge y \mapsto 1}_e$, %
\fi{}%
which clearly does not hold, since executing $c$ on a state satisfying the
precondition leads to a successful final state mapping $x$ to $1$.

For the frame rule to be recovered, it needs to take errors into account.  The
solution lies on the reachability properties of memory safety: instead of
enforcing locality by preventing errors, we can use the fact that memory
operations in a safe language are automatically local---in particular, local to
the identifiers that the program possesses.

\begin{theorem}
  \label{thm:weak-frame-rule}
  Under the same assumptions as \Cref{thm:frame-rule}, the following rule is
  sound
  \[
  \inferrule*[Right=SafeFrame]
  { \independent(r, \modvars(c)) \and \triple{p}{c}{q}_e }
  { \triple{p \triangleright r}{c}{q \triangleright r}_e }
  \]
  where $p \mathrel{\triangleright} r$ denotes the \emph{isolating conjunction}
  of $p$ and $r$, defined as %
  \ifpostsubmission{}%
  \[ \{ (l, m_1 \cup m_2) \mid (l, m_1) \in p, (l, m_2) \in r,
    \ids(l, m_1) \fresh \blocks(l, m_2) \}. \] %
  \fi{}%
\end{theorem}

The proof is similar to the one for the original rule, but it relies
additionally on \Cref{thm:frame-error}. This explains why the isolating
conjunction is needed, since it ensures that the fragment satisfying $r$ is
unreachable from the rest of the state.


\subsection{Discussion}

As hinted by their connection with the frame rule, the theorems of
\Cref{sec:frame-theorems} are a form of local reasoning: to reason about a
command, it suffices to consider its reachable state; \emph{how} this state is
used bears no effect on the unreachable portions.  In a more realistic language,
reachability might be inferred from additional information such as typing.  But
even here it can probably be accomplished by a simple check of the program text.

For example, consider the hypothetical jpeg decoder from
\Cref{sec:introduction}. We would like to guarantee that the decoder cannot
tamper with an unreachable object---a window object, a whitelist of trusted
websites, etc. The frame theorems give us a means to do so, provided that we are
able to show that the object is indeed unreachable; additionally, they imply
that the jpeg decoder cannot directly extract any information from this
unreachable object, such as passwords or private keys.

Many real-world attacks involve direct violations of these reasoning principles.
For example, consider the infamous Heartbleed attack on OpenSSL, which used
out-of-bounds reads from a buffer to leak data from completely unrelated parts
of the program state and to steal sensitive
information~\cite{DurumericKAHBLWABPP14}.  Given that the code fragment that
enabled that attack was just manipulating an innocuous array, a programmer could
easily be fooled into believing (as probably many have) that that snippet could
not possibly access sensitive information, allowing that vulnerability to remain
unnoticed for years.

Finally, our new frame rule only captures the fact that a command cannot
influence the heap locations that it cannot reach, while our noninterference
result (\Cref{cor:noninterference}) captures not just this integrity aspect of
memory safety, but also a secrecy aspect.
We hope that future research will explore the connection between the secrecy
aspect of memory safety and (relational) program logics.


\iflater
\ch{BTW. Is there any relation between memory safety and Francois' Pottier
  anti-frame rule?}
\fi

\section{Relaxing Memory Safety}
\label{sec:relaxations}

So much for formalism.  What about reality?
\ifpostsubmission{}\fi{}%
Strictly speaking, the security properties we have identified do not hold of any
real system.  This is partly due to fundamental physical limitations---real
systems run with finite memory, and interact with users in various ways that
transcend inputs and outputs, notably through time and other side
channels.\ifpostsubmission\footnote{%
  Though the attacker model considered in this paper does not try to address
  such side-channel attacks, one should be able to use the previous research on
  the subject to protect against them or limit the damage they can cause
  \cite{Smith09, BackesKR09, ZhangAM12, StefanBYLTRM13}.} \fi{}%
A more interesting reason is that real systems typically do not impose all the
restrictions required for the proofs of these properties.  Languages that aim
for safety generally offer relatively benign glimpses of their implementation
details (such accessing the contents of uninitialized memory, extract physical
addresses from pointers or compare them for ordering) in return for significant
flexibility or performance gains.
In other systems, the concessions are more fundamental, to the extent that it is
harder to clearly delimit what part of a program is unsafe: the SoftBound
transformation~\cite{NagarakatteZMZ09}, for example, adds bounds checks for C
programs, but does not protect against memory-management bugs; a related
transformation, CETS~\cite{NagarakatteZMZ10}, is required for temporal
safety. \iflater\bcp{I think we have to let it go for now, but even after
  rewriting I still find this paragraph confused.}\fi

In this section, we enumerate common relaxed models of memory safety and
evaluate how they affect the reasoning principles and security guarantees of
\Cref{sec:reasoning}. Some relaxations, such as allowing pointers to be forged
out of thin air, completely give up on reachability-based reasoning.  Others,
however, retain strong guarantees for integrity while giving up on some secrecy,
allowing aspects of the global state of a program to be observed. For example, a
system with finite memory (\Cref{sec:infinite-memory}) may leak some information
about its memory consumption, and a system that allows pointer-to-integer casts
(\Cref{sec:observing-pointers}) may leak information about its memory layout. %
Naturally, the distinction between integrity and secrecy should be taken with a
grain of salt, since the former often depends on the latter; for example, if a
system grants privileges to access some component when given with the right
password, a secrecy violation can escalate to an integrity violation!

\subsection{Forging Pointers}
\label{sec:forging-pointers}

Many real-world C programs \iffull\else{}use \fi{}integers as
pointers.  If this idiom is allowed without restrictions, then {}local reasoning is compromised, as every memory region may be reached from
anywhere in the program.
It is not surprising that languages that strive for {}safety
either forbid this kind of pointer forging or confine it to
\ifpostsubmission{}clear\fi{} unsafe fragments.

More insidiously, and perhaps surprisingly, similar dangers 
lurk in the stateful abstractions of some systems that are widely regarded as
``memory safe.''  JavaScript, for example, allows code to access
\emph{arbitrary} global variables by indexing an associative array with a
string, a feature that enables many serious attacks~\cite{caja, FournetSCDSL13,
  TalyEMMN11, MeyerovichL10}.  One might argue that global variables in
JavaScript are ``memory unsafe'' because they fail to validate local reasoning:
even if part of a JavaScript program does not explicitly mention a given global
variable, it might still change this variable or the objects it points
to. Re-enabling local reasoning requires strong restrictions on the programming
style~\cite{BhargavanDM13, caja, FournetSCDSL13}.

\subsection{Observing Pointers}
\label{sec:observing-pointers}

The language of \Cref{sec:imp} maintains a complete separation between pointers
and other values.  In reality, this separation is often only enforced in one
direction.  For example, some tools for enforcing memory safety in
C~\cite{NagarakatteZMZ09, DeviettiBMZ08} allow pointer-to-integer casts
\cite{KangHMGZV15} (a feature required by many low-level
idioms~\cite{cheri_asplos2015, MemarianMLNCWS16}); and the default
implementation of \verb!hashCode()! in Java leaks address information.
To model such features, we can extend the syntax of expressions with a form
$\cast(e)$, \iflater\bcp{Since ``cast'' is usually a more general thing, perhaps
  we can choose a name that explicitly signals that we are exactly casting
  pointers to integers?}\ch{to-int? addr-of?}\fi{} the semantics of which are
defined with some function $\lsb \cast\rsb : \I\times \Z \to \Z$ for converting
a pointer to an integer:
\begin{align*}
  \lsb\cast(e)\rsb(s) & = \lsb\cast\rsb(\lsb e\rsb(s)) \qquad \text{ if $\lsb
    e\rsb(s) \in \I \times \Z$}
\end{align*}

Note that the original language included an operator for extracting the offset
of a pointer.  Their definitions are similar, but have crucially different
consequences: while offsets do not depend on the identifier, allocation order,
or other low-level details of the language implementation (such as the choice of
physical addresses when allocating a block), all of these could be relevant when
defining the semantics of $\cast$.  The three frame theorems
(\ref{thm:frame-ok}, \ref{thm:frame-loop}, and \ref{thm:frame-error}) are thus
lost, because the state of unreachable parts of the heap may influence integers
observed by the program. An important consequence is that secrecy is weakened in
this language: an attacker could exploit pointers as a side-channel to learn
secrets about data it shouldn't access.

Nevertheless, \emph{integrity} is not affected: if a block is unreachable, its
contents will not change at the end of the execution. (This result was also
proved in Coq.)
\begin{theorem}[Integrity-only Noninterference]
  \label{thm:integrity-noninterference}
  Let $s_1$, $s_2$, and $s'$ be states and $c$ a command such that $\vars(c)
  \subseteq \vars(s_1)$, $\ids(s_1) \fresh \blocks(s_2)$, and $\lsb c\rsb(s_1
  \cup s_2) = s'$.  Then we can find $s_1' \in \St$ such that $s' = s_1' \cup
  s_2$ and $\ids(s_1') \fresh \blocks(s_2)$.
\end{theorem}

\iflater
\ch{Q: Is this result enough to obtain the frame rule. Intuitively your frame
  rule should still hold, since it only needs integrity, not secrecy.}
\aaa{That is a good point. This result is not enough to obtain the frame rule.
  However, I do believe that it should be possible to prove soundness for
  both frame rules in this setting.}
\fi

\iflater
\aaa{Need to ensure this name matches the one in the Coq development}
\fi

The stronger noninterference result of \Cref{cor:noninterference} showed that,
if pointer-to-integer casts are prohibited, changing the contents of the
unreachable portion $s_2$ has no effect on the reachable portion, $s_1'$.
In contrast, Theorem \ref{thm:integrity-noninterference} allows changes in $s_2$
to influence $s_1'$ in arbitrary ways in the presence of these casts: not only
can the contents of this final state change, but the execution can also loop
forever or terminate in an error.

To see why, suppose that the jpeg decoder of \Cref{sec:introduction} is part of
a web browser, but that it does not have the required pointers to learn the
address that the user is currently visiting.  Suppose that there is some
relation between the memory consumption of the program and that website, and
that there is some correlation between the memory consumption and the identifier
assigned to a new block.  Then, by allocating a block and converting its pointer
to a integer, the decoder might be able to infer useful information about the
visited website \cite{JanaS12a}.  Thus, if $s_2$ denoted the part of the state
where that location is stored, changing its contents would have a nontrivial
effect on $s_1'$, the part of the state that the decoder does have access to.
We could speculate that, in a reasonable system, this channel can only reveal
information about the layout of unreachable regions, and not their
contents. Indeed, we conjecture this for the \ifpostsubmission language of \fi this subsection.
\iflater
\ch{It would strengthen this section to actually do this}
\fi

Finally, it is worth noting that simply excluding casts might not suffice to
prevent this sort of vulnerability.
Recall that our language takes both offsets and identifiers into account for
equality tests.  For performance reasons, we could have chosen a different
design that only compares physical addresses, completely discarding identifiers.
If attackers know the address of a pointer in the program---which could happen,
for instance, if they have access to the code of the program and of the
allocator---they can use pointer arithmetic (which is generally harmless and
allowed in our language) to find the address of other pointers.  If $x$ holds
the pointer they control, they can run, for instance,
\[ y \gets \calloc(1); \cifte{x + 1729 = y}{\ldots}{\ldots}, \] to learn the
location assigned to $y$ and draw conclusions about the global state.

\subsection{Uninitialized Memory}
\label{sec:uninitialized}

Safe languages typically initialize new variables and objects.  But this can
degrade performance, leading to cases where this feature is dropped---including
standard C implementations, safer alternatives~\cite{NagarakatteZMZ09,
  DeviettiBMZ08}, OCaml's \texttt{Bytes.create} primitive, or Node.js's
\texttt{Buffer.allocUnsafe}, for example.

The problem with this concession is that the entire memory becomes relevant to
execution, and local reasoning becomes much harder. By inspecting old values
living in uninitialized memory, an attacker can learn about parts of the state
they shouldn't access and violate secrecy.  This issue would become even more
severe in a system that allowed old pointers or other capabilities to occur in
re-allocated memory in a way that the program can use, since they could yield
access to restricted resources directly, leading to potential integrity
violations as well. (The two examples given above---OCaml and Node.js---do not
suffer from this issue, because any preexisting pointers in re-allocated memory
are treated as bare bytes that cannot be used to access memory.)

\iflater
\aaa{TODO: Look for a system that allows this kind of capability leak.}
\fi

\subsection{Dangling Pointers and Freshness}
\label{sec:freshness}

Another crucial issue is the treatment of dangling pointers---references to
previously freed objects.  Dangling pointers are problematic because there is an
inherent tension between giving them a sensible semantics (for instance, one
that validates the properties of \Cref{sec:reasoning}) and obtaining good
performance and predictability.  Languages with garbage collection avoid the
issue by forbidding dangling pointers altogether---heap storage is freed only
when it is unreachable.  In the language of \cref{sec:imp}, besides giving a
well-defined behavior to the use of dangling pointers (signaling an error), we
imposed strong freshness requirements on allocation, mandating not only that the
new identifier not correspond to any existing block, but also that it not be
present {\em anywhere else} in the state.

To see how the results of~\Cref{sec:reasoning} are affected by weakening
freshness, suppose we run the program \ifpostsubmission{}%
$x \gets \calloc(1); z \gets (y = x)$ \fi{}%
on a state where $y$ holds a dangling pointer. Depending on the allocator and
the state of the memory, the pointer assigned to $x$ could be equal to
$y$. Since this outcome depends on the entire state of the system, not just the
reachable memory, \Cref{thm:frame-ok,thm:frame-loop,thm:frame-error} now
fail. Furthermore, an attacker with detailed knowledge of the allocator could
launder secret information by testing pointers for equality.  Weakening
freshness can also have integrity implications, since it becomes harder to
ensure that blocks are properly isolated. For instance, a newly allocated block
might be reachable through a dangling pointer controlled by an attacker,
allowing them to access that block even if they were not supposed to.

Some practical solutions for memory safety use mechanisms similar to our
language's, where each memory location is tagged with an identifier describing
the region it belongs to~\cite{ClauseDOP07,pump_asplos2015}.  Pointers are
tagged similarly, and when a pointer is used to access memory, a violation is
detected if its identifier does not match the location's.  However, for
performance reasons, the number of possible identifiers might be limited to a
relatively small number, such as 2 or 4~\cite{ClauseDOP07} or 16~\cite{m7negative}.
In addition to the problems above, since multiple live regions can share the same
identifier in such schemes, it might be possible for buffer overflows to lead to
violations of secrecy and integrity as well.

Although we framed our discussion in terms of identifiers, the issue of
freshness can manifest itself in other ways.  For example, many systems for
spatial safety work by adding base and bounds information to pointers.  In some
of these~\cite{DeviettiBMZ08,NagarakatteZMZ09}, dangling pointers are treated as
an orthogonal issue, and it is possible for the allocator to return a new memory
region that overlaps with the range of a dangling pointer, in which case the new
region will not be properly isolated from the rest of the state.

Finally, dangling pointers can have disastrous consequences for overall
system security, independently of the freshness issues just described: freeing a
pointer more than once can break allocator invariants, enabling attacks~\cite{Szekeres2013}.

\subsection{Infinite Memory}
\label{sec:infinite-memory}

Our idealized language allows memory to grow indefinitely.  But real languages
run on finite memory, and allocation fails when programs run out of space.
Besides enabling denial-of-service attacks, finite memory has consequences for
secrecy.  \Cref{cor:noninterference} does not hold in a real programming
language as is, because an increase in memory consumption can cause a previously
successful allocation to fail.  By noticing this difference, a piece of code
might learn something about the \emph{entire} state of the program.  How
problematic this is in practice will depend on the particular system under
consideration.

A potential solution is to force programs that run out of memory to terminate
immediately.  Though this choice might be bad from an availability standpoint,
it is probably the most benign in terms of secrecy.  We should be able to prove
an \emph{error-insensitive} variant of \Cref{cor:noninterference}, where the
only significant effect that unreachable memory can have is to turn a successful
execution or infinite loop into an error.
\iflater
\ch{It would strengthen this section to actually do this}
\fi
Similar issues arise for \iffull\else IFC \fi
mechanisms that often cannot prevent secrets from influencing program
termination, leading to \emph{termination-insensitive} notions of
noninterference.

Unfortunately, even an error-insensitive result might be too strong for real
systems, which often make it possible for attackers to extract multiple bits of
information about the global state of the program---as previously noted in the
IFC literature~\cite{askarov08:tini_leaks_more_than_1_bit}. Java, for example,
does not force termination when memory runs out, but triggers an exception that
can be caught and handled by user code, which is then free to record the event
and probe the allocator with a different test.  And most languages do not
operate in batch mode like ours does, merely producing a single answer at the
end of execution; rather, their programs continuously interact with their
environment through inputs and outputs, allowing them to communicate the exact
amount of memory that caused an error.
\iflater
\ch{It can
  get worse than this if out of memory is not properly detected.  I vaguely
  remember C programs that segfault when running out of memory. \ch{It seems
  a Linux thing but not really a security issue}  We could also
  check whether the C standard calls this an undefined behavior.
  \ch{Unclear, leaving this for later
\url{https://cansecwest.com/core05/memory_vulns_delalleau.pdf}}}
\fi

This discussion suggests that, if size vulnerabilities are a real concern, they
need to be treated with special care. One approach would be to limit the amount
of memory an untrusted component can allocate\iffull
\else~\cite{Yang:2014}\fi, so that exhausting
the memory allotted to that component doesn't reveal information about the state
of the rest of the system (and so that also global denial-of-service attacks are
prevented).  A more speculative idea is to develop \emph{quantitative}
versions~\cite{Smith09, BackesKR09} of the noninterference results discussed
here that apply only if the total memory used by the program is below a certain
limit.

\ifpostsubmission{}
\fi{}

\section{Case Study: A Memory-safety Monitor}
\label{sec:micro-policy}

To demonstrate the applicability of our characterization, we use it to analyze a
tag-based monitor proposed by Dhawan \ETAL to enforce heap safety for low-level
code~\cite{pump_asplos2015}.  In prior work~\cite{micropolicies2015}, we and
others showed that an idealized model of the monitor correctly implements a
higher-level abstract machine with built-in memory safety---a bit more formally,
every behavior of the monitor is also a behavior of the abstract machine. %
Building upon this work, we prove that this abstract machine satisfies a
noninterference property similar to \Cref{cor:noninterference}.  We were also
able to prove that a similar result holds for a lower-level machine that runs a
so-called ``symbolic'' representation of the monitor---although we had to
slightly weaken the result to account for memory exhaustion
(cf. \Cref{sec:infinite-memory}), since the machine that runs the monitor has
finite memory, while the abstract machine has infinite memory.  If we had a
verified machine-code implementation of this monitor, it would be possible to
prove a similar result for it as well.

\subsection{Tag-based Monitor}
\label{sec:monitor}

We content ourselves with a brief overview of Dhawan~\ETAL's
monitor~\cite{pump_asplos2015,micropolicies2015}, since the formal statement of
the reasoning principles it supports are more complex than the one for the
abstract machine from \cref{sec:abstract}, on which we will focus.
\ifpostsubmission{}\fi{}%
Following a proposal by Clause \emph{et al.}~\cite{ClauseDOP07}, Dhawan~\ETAL's
monitor enforces memory safety for heap-allocated data by checking and
propagating \emph{metadata tags}.  Every memory location receives a tag that
uniquely identifies the allocated region to which that location belongs (akin to
the identifiers in \Cref{sec:imp}), and pointers receive the tag of the region
they are allowed to reference.  The monitor assigns these tags to new regions by
storing a monotonic counter in protected memory that is bumped on every call to
\verb!malloc!; with a large number of possible tags, it is possible to avoid the
freshness pitfalls discussed in \Cref{sec:freshness}.  When a memory access
occurs, the monitor checks whether the tag on the pointer matches the tag on the
location.  If they do, the operation is allowed; otherwise, execution halts.
The monitor instruments the allocator to make set up tags correctly.  Its
implementation achieves good performance using the \emph{PUMP}, a hardware
extension accelerating such micro-policies for metadata tagging
\cite{pump_asplos2015}.

\subsection{Abstract Machine}
\label{sec:abstract}

\newcommand{\rs}[0]{\mathit{rs}}
\newcommand{\pc}[0]{\mathit{pc}}

The memory-safe abstract machine~\cite{micropolicies2015} operates on two kinds
of values: machine words $w$, or pointers $(i, w)$, which are pairs of an
identifier $i \in \I$ and an offset $w$.  We use $\W$ to denote the set of
machine words, and $\V$ to denote the set of values. Machine states are triples
$(m, \rs, \pc)$, where \ifpostsubmission{}%
(1) $m \in \I \partfunfin \V^*$ is a \emph{memory} mapping identifiers to lists
of values; (2) $\rs \in \R \partfunfin \V$ is a \emph{register bank}, mapping
register names to values; and %
(3) $\pc \in \V$ is the \emph{program counter}.  \fi{}


\ifpostsubmission{}%
\fi{}%

The execution of an instruction is specified by a step relation $s \to s'$.  If
there is no $s'$ such that $s \to s'$, we say that $s$ is stuck, which means
that a fatal error occurred during execution.  On each instruction, the machine
checks if the current program counter is a pointer and, if so, tries to fetch
the corresponding value in memory.  The machine then ensures that this value is
a word that correctly encodes an instruction and, if so, acts accordingly.
\ifpostsubmission{}%
The instructions of the machine, representative of typical RISC architectures,
allow programs to perform binary and logical operations, move values to and from
memory, and branch.  \fi{}%
The machine is in fact fairly similar to the language of \Cref{sec:imp}. Some
operations are overloaded to manipulate pointers; for example, adding a pointer
to a word is allowed, and the result is obtained by adjusting the pointer's
offset accordingly.  Accessing memory causes the machine to halt when the
corresponding position is undefined.

In addition to these basic instructions, the machine possesses a set of special
\emph{monitor services} that can be invoked as regular functions, using
registers to pass in arguments and return values.  There are two services
$\calloc$ and $\cfree$ for managing memory, and one service $\mathsf{eq}$ for
testing whether two values are equal.
The reason for using separate monitor services instead of special instructions is
to keep its semantics closer to the more concrete machine that implements it.
While instructions include an equality test, it cannot replace the $\mathsf{eq}$
service, since it only takes physical addresses into account.
As argued in \Cref{sec:observing-pointers}, such comparisons can be turned into
a side channel\ifpostsubmission{}\fi{}.  To prevent this, testing two pointers for
equality directly using the corresponding machine instruction results in an
error if the pointers have different block identifiers.

\subsection{Verifying Memory Safety}

The proof of memory safety for this abstract machine mimics the one carried for
the language in \Cref{sec:reasoning}.  We use similar notations as before: $\pi
\cdot s$ means renaming every identifier that appears in $s$ according to the
permutation $\pi$, and $\ids(s)$ is the finite set of all identifiers that
appear in the state $s$.
A simple case analysis on the possible instructions yields analogs of
\Cref{thm:renaming,thm:frame-ok,thm:frame-error} (we don't include an analog of
\Cref{thm:frame-loop} because we consider individual execution steps, where
loops cannot occur).  {}

\begin{theorem}
  \label{thm:mp-renaming}
  Let $\pi$ be a permutation, and $s$ and $s'$ be two machine states such that
  $s \to s'$.  There exists another permutation $\pi'$ such that $\pi \cdot s
  \to \pi' \cdot s'$.
\end{theorem}

\begin{theorem}
  \label{thm:mp-frame-ok}
  Let $(m_1, \rs, \pc)$ be a state of the abstract machine, and $m_2$ a
  memory. Suppose that $\ids(m_1, \rs, \pc) \fresh \dom(m_2)$, and that $(m_1, \rs,
  \pc) \to (m', \rs', \pc')$.  Then, there exists a permutation $\pi$ such that
  \ifpostsubmission{}%
  $\ids(\pi \cdot m', \pi \cdot \rs, \pi \cdot \pc) \fresh \dom(m_2)$ and
  $(m_2 \cup m_1, \rs, \pc) \to (m_2 \cup \pi \cdot m', \pi \cdot \rs', \pi \cdot
  \pc')$.
  \fi{}%
\end{theorem}

\begin{theorem}
  \label{thm:mp-frame-error}
  Let $(m_1, \rs, \pc)$ be a machine state, and $m_2$ a memory. If
  $\ids(m_1, \rs, \pc) \fresh \dom(m_2)$, and $(m_1, \rs, \pc)$ is stuck, then
  $(m_2 \cup m_1, \rs, \pc)$ is also stuck.
\end{theorem}

\ifpostsubmission{}%
Once again, we can combine these properties to obtain a proof of
noninterference.  Our Coq development includes a complete statement.
\fi{}%

\subsection{Discussion}

The reasoning principles supported by the memory-safety monitor have an
important difference compared to the ones of \Cref{sec:reasoning}.  In the
memory-safe language, reachability is relative to a program's local
variables. If we want to argue that part of the state is isolated from some code
fragment, we just have to consider that fragment's local variables---other parts
of the program are still allowed to access the region.  The memory-safety
monitor, on the other hand, does not have an analogous notion: an unreachable
memory region is useless, since it remains unreachable by all components
forever.

It seems that, from the standpoint of noninterference, heap memory safety
\emph{taken in isolation} is much weaker than the guarantees it provides in the
presence of other language features, such as local variables. Nevertheless, the
properties studied above suggest several avenues for strengthening the mechanism
and making its guarantees more useful.  The most obvious one would be to use the
mechanism as the target of a compiler for a programming language that provides
other (safe) stateful abstractions, such as variables and a stack for procedure
calls.  A more modest approach {}would be to add other state abstractions to the mechanism
itself.  Besides variables and call stacks, if the mechanism made code immutable
and separate from data, a simple check would suffice to tell whether a code
segment stored in memory references a given privileged register.  If the
register is the only means of reaching a memory region, we should be able to
soundly infer that that code segment is independent of that region.
%

On a last note, although the abstract machine we verified is fairly close to our
original language, the dynamic monitor that implements it using tags is quite
different (\cref{sec:monitor}).  In particular, the monitor works on a machine
that has a flat memory model, and keeps track of free and allocated memory using
a protected data structure that stores block metadata.  It was claimed that
reasoning about this base and bounds information was the most challenging part
of the proof that the monitor implements the abstract
machine~\cite{micropolicies2015}.  For this reason, we believe that this proof
can be adapted to other enforcement mechanisms that rely solely on base and
bounds information---for example, fat pointers~\cite{LowFat2013,DeviettiBMZ08}
or SoftBound~\cite{NagarakatteZMZ09}---while keeping a similar abstract machine
as their specification, and thus satisfying a similar noninterference
property. This gives us confidence that our memory safety characterization
generalizes to other settings.

\section{Related Work}
\label{sec:related-work}

The present work lies at the intersection of two areas of previous research:
one on formal characterizations of memory safety, the other on reasoning
principles for programs.  We review the most closely related work in these
areas.

\paragraph*{Characterizing Memory Safety}

Many formal characterizations of memory safety originated in attempts to
reconcile its benefits with low-level code.  Generally, these works claim that a
mechanism is safe by showing that it prevents or catches typical temporal and
spatial violations.  Examples in the literature include:
Cyclone~\cite{SwamyHMGJ06}, a language with a {}type
system for safe manual memory management; %
CCured~\cite{ccured_toplas2005}, a program transformation that adds temporal
safety to C by refining its pointer type\iffull\else{}
with\fi{} various degrees of safety; %
Ivory~\cite{ElliottPWHBSSL15} an embedding of a similar ``safe-C variant'' into
Haskell; %
SoftBound~\cite{NagarakatteZMZ09}, an instrumentation technique for C programs
for spatial safety, including the detection of bounds violations within an
object; %
CETS~\cite{NagarakatteZMZ10}, a compiler pass for preventing temporal safety
violations in C programs, including accessing dangling pointers into freed heap
regions and stale stack frames; %
the memory-safety monitor for the PUMP~\cite{pump_asplos2015,micropolicies2015},
which formed the basis of our case study in \Cref{sec:micro-policy}; and %
languages like Mezzo~\cite{pottier-protzenko-13} and Rust~\cite{Turon17}, whose
guarantees extend to preventing data races~\cite{BalabonskiPP14}.  Similar
models appear in formalizations of C~\cite{LeroyB08,Krebbers15}, which need to
rigorously characterize its sources of undefined behavior---in particular,
instances of memory misuse.

Either explicitly or implicitly, these works define memory errors as attempts to
use a pointer to access a location that it was not meant to access---for
example, an out-of-bounds or free one.  This was noted by
Hicks~\cite{Hicks:memory-safety}, who, inspired by SoftBound, proposed to define
memory safety as an execution model that tracks what part of memory each pointer
can access.  Our characterization is complementary to these accounts, in that it
is \emph{extensional}: its data isolation properties allow us to reason directly
about the observable behavior of the program.  Furthermore, as demonstrated by our
application to the monitor of \Cref{sec:micro-policy} and the discussions on
\Cref{sec:relaxations}, it can be adapted to various enforcement mechanisms and
variations of memory safety.


\paragraph*{Reasoning Principles}

Separation logic~\cite{Reynolds:2002,Yang:2002} has been an important source of
inspiration for our work.  The logic's frame rule enables its local reasoning
capabilities and imposes restrictions that are similar to those mandated by
memory-safe programming guidelines.  As discussed in \Cref{sec:separation}, our
reasoning principles are reminiscent of the frame rule, but use reachability to
guarantee locality in settings where memory safety is enforced automatically.
In separation logic, by contrast, locality needs to be guaranteed for each
program individually by comprehensive
proofs.

Several works have investigated similar reasoning principles for a variety of
program analyses, including static, dynamic, manual, or a mixture of those.
Some of these are formulated as expressive logical relations, guaranteeing that
programs are compatible with the framing of state invariants; representative
works include: L\({}^{\mbox{\small 3}}\)~\cite{AhmedFM07}, a linear calculus
featuring strong updates and aliasing control; the work of Benton and
Tabereau~\cite{Benton:2009} on a compiler for a higher-order language; and the
work of Devriese \emph{et al.}~\cite{DevriesePB16} on object capabilities for a
JavaScript-like language.  Other developments are based on proof systems
reminiscent of separation logic{}; these include Yarra~\cite{SchlesingerPSWZ14}, an extension of C
that allows programmers to protect the integrity of data structures marked as
\emph{critical}; the work of Agten \emph{et al.}~\cite{Agten0P15}, which allows
mixing unverified and verified components by instrumenting the program to check
that required assertions hold at interfaces; and the logic of Swasey \emph{et
  al.}~\cite{SwaseyGD17} for reasoning about object capabilities.

Unlike our work, these developments do not propose reachability-based isolation
as a general \emph{definition} of memory safety, nor do they attempt to analyze
how their reasoning principles are affected by common variants of memory safety.
Furthermore, many of these other works---especially the logical relations---rely
on encapsulation mechanisms such as closures, objects, or modules that go beyond
plain memory safety.  Memory safety alone can only provide complete isolation,
while encapsulation provides finer control, allowing some interaction between
components, while guaranteeing the preservation of certain state invariants. In
this sense, one can see memory-safety reasoning as a special case of
encapsulation reasoning.
Nevertheless, it is a practically relevant special case that is interesting on
its own, since when reasoning about an encapsulated component, one must argue
explicitly that the invariants of interest are preserved by the private
operations of that component; memory safety, on the other hand, guarantees that
\emph{any} invariant on unreachable parts of the memory is automatically
preserved.

Perhaps closer to our work, Maffeis \emph{et al.}~\cite{MaffeisMT10} show that
their notion of ``authority safety'' guarantees isolation, in the sense that a
component's actions cannot influence the actions of another component with
disjoint authority. Their notion of authority behaves similarly to the set of
block identifiers accessible by a program in our language; however, they do not
attempt to connect their notion of isolation to the frame rule, noninterference,
or traditional notions of memory safety.

Morrisett \emph{et al.}~\cite{Morrisett:1995} state a correctness criterion for
garbage collection based on program equivalence.  Some of the properties they
study are similar to the frame rule, describing the behavior of code running in
an extended heap.  However, they use this analysis to justify the validity of
deallocating objects, rather than studying the possible interactions between the
extra state and the program in terms of integrity and secrecy.

\iflater
\aaa{I'll have a look at the CSF paper
  (\href{https://people.mpi-sws.org/~dg/papers/csf17-hyperproperties.pdf}{Secure
    compilation and hyperproperty preservation})
  that you mentioned about full
  abstraction; I wonder if we can make a clear argument that full abstraction
  does not preserve that kind of reasoning!}
\fi






\iflater
\aaa{David \cite{SwaseyGD17} mentioned the (previously proposed) concept of
  robust safety in their work, which guarantees that programs have meaningful
  safety guarantees even when composed with untrusted code.  We should probably
  mention that somewhere.}
\fi



\iflater
\aaa{Need a pass here later}

Mezzo~\cite{pottier-protzenko-13} is a concurrent dialect of ML which rules out
errors such as data races while enabling certain idioms that are usually not
possible in a purely functional setting, such as strong updates (changing the
\emph{type} of a reference when assigning to it) and gradual, stateful
initialization of immutable data structures. Thus, it provides some guarantees
that are beyond what is usually seen as defining memory safety. The soundness
proofs for a fragment of the type system~\cite{BalabonskiPP14} include a
progress and preservation result, as well as the absence of data races, in the
sense that every piece of data can only be written by at most one thread
concurrently.  \iflater\bcp{There is a LOT of other formal work on formalizing
  object ownership systems in the OOPSLA / FOOL community.}\fi \fi

\iflater
\ch{One point that comes out in \\ \href{the discussion to Mike's
    post}{http://www.pl-enthusiast.net/2014/07/21/memory-safety/\#comment-453} is
  the difficulty to define memory safety for something that's as flexible as
  C. For this the Cambridge folks have some nice empirical
  studies~\cite{cheri_asplos2015}.}
\fi

\section{Conclusions and Future Work}
\label{sec:conclusion}

We have explored the consequences of memory safety for reasoning about programs,
formalizing intuitive principles that, we argue, capture the essential
distinction between memory-safe systems and memory-unsafe ones.  We showed how
the reasoning principles we identified apply to a recent dynamic monitor for
heap memory safety.

The systems studied in this paper have a simple storage model: the language of
\Cref{sec:imp} has just global variables and flat, heap-allocated arrays, while
the monitor of \Cref{sec:micro-policy} doesn't even have variables or immutable
code. Realistic programming platforms, of course, offer much richer stateful
abstractions, including, for example, procedures with stack-allocated local
variables as well as structured objects with contiguously allocated
sub-objects. In terms of memory safety, these systems have a richer vocabulary
for describing resources that programs can access, and programmers could benefit
from isolation-based local reasoning involving these resources.

For example, in typical safe languages with procedures, the behavior of a
procedure should depend only on its arguments, the global variables it uses, and
the portions of the state that are reachable from these values; if the caller of
that procedure has a private object that is not passed as an argument, it should
not affect or be affected by the call. Additionally, languages such as C allow
for objects consisting of contiguously allocated sub-objects for improved
performance. Some systems for spatial safety~\cite{NagarakatteZMZ09,
  DeviettiBMZ08} allow \emph{capability downgrading}---that is, narrowing the
range of a pointer so that it can't access outside of a sub-object's bounds. It
would be interesting to refine our model to take these features into account.
In the case of the monitor of \Cref{sec:micro-policy}, such considerations could
lead to improved designs or to the integration of the monitor inside a secure
compiler.
Conversely, it would be interesting to derive finer security properties for
relaxations{} like the ones discussed
in~\Cref{sec:relaxations}.  Some inspiration could come from the
\iffull\else{}IFC\fi{} literature, where quantitative
noninterference results provide bounds on the probability that some secret is
leaked, the rate at which it is leaked, how many bits are leaked, etc.
\cite{BackesKR09, Smith09}.

The main goal of this work was to understand, formally, the benefits of memory
safety for informal and partial reasoning, and to evaluate a variety of weakened
forms of memory safety in terms of which reasoning principles they preserve.
However, our approach may also suggest ways to improve program verification.
One promising idea is to leverage the
guarantees of memory safety to obtain proofs of program correctness
modulo unverified code that could have errors, in contexts where complete
verification is too expensive or not possible (\EG for programs with a
plugin mechanism).

\ifanon\else
\paragraph*{Acknowledgments}
We are grateful to
  Antal Spector-Zabusky,
  Greg Morrisett,
  Justin Hsu,
  Michael Hicks,
  Nick Benton,
  Yannis Juglaret,
  William Mansky,
and
  Andrew Tolmach
for useful suggestions on earlier drafts.
\bcp{add a bunch of people that we had email discussions with...}
This work is supported by NSF grants Micro-Policies (1513854) and DeepSpec
(1521523), DARPA SSITH/HOPE, and ERC Starting Grant SECOMP (715753).
\fi

\iflater
\asz{Careful with your bibliography -- you have some empty fields (producing
  output like ``\ldots ISSN 0362-1340. . URL\ldots'') and you should be
  consistent with your URLS (perhaps have them all be
  \url{http://dx.doi.org/...} as far as possible, and avoid
  preprints/abstracts/etc.?).}\ch{Preprints are great. Paywalls suck.}
\fi

\clearpage
\appendix
\section*{Appendix}

\begin{figure}
  \[
  \begin{array}{rclr}
    \oplus & ::=  & {+} \mid {\times} \mid {-} \mid {=} \mid {\leq} \mid {\cand} \mid {\cor} & \text{(operators)}\\
    e      & ::=  & x \in \var \mid b \in \B \mid n \in \Z & \text{(expressions)} \\
           & \mid & e_1 \oplus e_2 \mid \cnot\,e \mid \coffset\,e \mid \nil & \\
    c      & ::=  & \cskip \mid c_1; c_2 & \text{(commands)}\\
           & \mid & \cifte{e}{c_1}{c_2} & \\
           & \mid & \cwhiledo{e}{c} & \\
           & \mid & x \gets e \mid x \gets [e] \mid [e_1] \gets e_2 & \\
           & \mid & x \gets \calloc(e) \mid \cfree(e)
  \end{array}
  \]
\smallskip
  \statetablelong
  \caption{Syntax and program states}
  \label{fig:formal-syntax}
\end{figure}

\begin{figure}
{\footnotesize
\begin{alignat*}{2}
  \lsb x\rsb(l, m)
  &&&\teq
  \begin{cases}
    l(x) & \text{if $x \in \dom(l)$} \\
    \nil & \text{otherwise} \\
  \end{cases} \\
  \lsb b\rsb(s) &&&\teq b \\
  \lsb n\rsb(s) &&&\teq n \\
  \lsb \nil\rsb(s) &&&\teq \nil \\
  \lsb e_1 + e_2\rsb(s)
  &&&\teq
  \begin{cases}
    n_1 + n_2
    & \text{if $\lsb e_1\rsb(s) = n_1$ and $\lsb e_2\rsb(s) = n_2$} \\
    (i, n_1 + n_2)
    & \text{if $\lsb e_1\rsb(s) = (i, n_1)$ and $\lsb e_2\rsb(s) = n_2$} \\
    & \text{or $\lsb e_1\rsb(s) = n_1$ and $\lsb e_2\rsb(s) = (i, n_2)$} \\
    \nil
    & \text{otherwise}
  \end{cases} \\
  \lsb e_1 - e_2\rsb(s)
  &&&\teq
  \begin{cases}
    n_1 - n_2
    & \text{if $\lsb e_1\rsb(s) = n_1$ and $\lsb e_2\rsb(s) = n_2$} \\
    (i, n_1 - n_2)
    & \text{if $\lsb e_1\rsb(s) = (i, n_1)$ and $\lsb e_2\rsb(s) = n_2$} \\
    \nil
    & \text{otherwise}
  \end{cases} \\
  \lsb e_1 \times e_2\rsb(s)
  &&&\teq
  \begin{cases}
    n_1 \times n_2
    & \text{if $\lsb e_1\rsb(s) = n_1$ and $\lsb e_2\rsb(s) = n_2$} \\
    \nil
    & \text{otherwise}
  \end{cases} \\
  \lsb e_1 = e_2\rsb(s)
  &&&\teq (\lsb e_1\rsb(s) = \lsb e_2\rsb(s)) \\
  \lsb e_1 \leq e_2\rsb(s)
  &&&\teq
  \begin{cases}
    n_1 \leq n_2
    & \text{if $\lsb e_1\rsb(s) = n_1$ and $\lsb e_2\rsb(s) = n_2$} \\
    \nil
    & \text{otherwise}
  \end{cases} \\
  \lsb e_1 \cand e_2\rsb(s)
  &&&\teq
  \begin{cases}
    b_1 \wedge b_2
    & \text{if $\lsb e_1\rsb(s) = b_1$ and $\lsb e_2\rsb(s) = b_2$} \\
    \nil
    & \text{otherwise}
  \end{cases} \\
  \lsb e_1 \cor e_2\rsb(s)
  &&&\teq
  \begin{cases}
    b_1 \vee b_2
    & \text{if $\lsb e_1\rsb(s) = b_1$ and $\lsb e_2\rsb(s) = b_2$} \\
    \nil
    & \text{otherwise}
  \end{cases} \\
  \lsb \cnot\;e\rsb(s)
  &&&\teq
  \begin{cases}
    \neg b
    & \text{if $\lsb e\rsb(s) = b$} \\
    \nil
    & \text{otherwise}
  \end{cases} \\
  \lsb \coffset\;e\rsb(s)
  &&&\teq
  \begin{cases}
    n
    & \text{if $\lsb e\rsb(s) = (i, n)$} \\
    \nil
    & \text{otherwise}
  \end{cases}
\end{alignat*}
}
\caption{Expression evaluation}
\label{fig:expression-sem}
\end{figure}

\begin{figure}
{\footnotesize
\centering
\begin{alignat*}{2}
     \bind(f, \bot) &&&\teq \bot
\\
     \bind(f, \oerror) &&&\teq \oerror
\\
     \bind(f, (I, l, m))
     &&&\teq
     \begin{cases}
       (I \cup I', l', m')
       & \text{if $f(l, m) = (I', l', m')$} \\
       \oerror
       & \text{if $f(l, m) = \oerror$} \\
       \bot
       & \text{otherwise}
     \end{cases}
\\
     \cif(b, x, y)
     &&&\teq
     \begin{cases}
       x
       & \text{if $b = \ctrue$} \\
       y
       & \text{if $b = \cfalse$} \\
       \oerror & \text{otherwise}
     \end{cases}
\end{alignat*}
}
\caption{Auxiliary operators $\bind$ and $\cif$}
\label{fig:bind}
\end{figure}

\begin{figure*}
  \begin{mathpar}
     \lsb \cskip\rsb_{+}(l, m) \teq (\emptyset, l, m)

     \lsb c_1; c_2\rsb_{+}(l, m)
     \teq \bind(\lsb c_2\rsb_{+}, \lsb c_1\rsb_{+}(l, m))

     \lsb \cifte{e}{c_1}{c_2}\rsb_{+}(l, m)
     \teq \cif(\lsb e\rsb(l, m), \lsb c_1 \rsb_{+}(l, m), \lsb c_2 \rsb_{+}(l, m))

     \lsb \cwhiledo{e}{c}\rsb_{+}
     \teq \codeface{fix}(\lambda\,f\,(l, m).\;
                        \cif(\lsb e\rsb(l, m),
                             \bind(\lsb c\rsb_{+}, f(l, m)),
                             (\emptyset, l, m)))

    \lsb x \gets e\rsb_{+}(l, m)
    \teq (\emptyset, l[x \mapsto \lsb e\rsb(l, m)], m)

    \lsb x \gets [e] \rsb_{+}(s)
    \teq
    \begin{cases}
      (\emptyset, l[x \mapsto v], m)
      & \text{if $\lsb e \rsb(s) = (i, n)$ and $m(i, n) = v$} \\
      \oerror
      & \text{otherwise}
    \end{cases}

    \lsb [e_1] \gets e_2 \rsb_{+}(s)
    \teq
    \begin{cases}
      (\emptyset, l, m[(i, n) \mapsto \lsb e_2\rsb(l, m)])
      & \text{if $\lsb e_1 \rsb(s) = (i, n)$ and $m(i, n) \neq \bot$} \\
      \oerror
      & \text{otherwise}
    \end{cases}

    \lsb x \gets \calloc(e)\rsb_{+}(l, m)
    \teq
    \begin{cases}
      (\{i\}, l[x \mapsto (i, 0)],
       m[(i, k) \mapsto 0 \mid 0 \leq k < n])
       & \text{if $\lsb e\rsb(l, m) = n$ and $i = \freshf(\ids(l, m))$} \\
       \oerror
       & \text{otherwise}
     \end{cases}

     \lsb \cfree(e)\rsb_{+}(l, m)
     \teq
     \begin{cases}
       (\emptyset, l, m[(i, k) \mapsto \bot \mid k \in \Z])
       & \text{if $\lsb e\rsb(l, m) = (i, 0)$ and $m(i, n) \neq \bot$ for some
         $n$} \\
       \oerror & \text{otherwise}
     \end{cases}
   \end{mathpar}

   \caption{Command evaluation with explicit allocation sets}
   \label{fig:command-sem}
\end{figure*}

\label{sec:semantics}

This appendix defines the language of \Cref{sec:imp} more formally.
\Cref{fig:formal-syntax} summarizes the syntax of programs and repeats the
definition of program states. The syntax is standard for a simple imperative
language with pointers.\iflater\ch{If we need space iffull to remove repetition}\fi

\Cref{fig:expression-sem} defines expression evaluation,
  $\lsb e \rsb : \St \to \V$.
Variables are looked up in the local-variable part of the state (for simplicity,
heap cells cannot be dereferenced in expressions; the command $x \gets [e]$ puts
the value of a heap cell in a local variable).  Constants (booleans, numbers,
and the special value $\nil$ used to simplify error propagation)
evaluate to themselves.
Addition and subtraction can be applied both to numbers and to combinations of
numbers and pointers (for pointer arithmetic); multiplication only works on
numbers.
Equality is allowed both on pointers and on numbers. Pointer equality compares
both the block identifier and its offset, and while this is harder to implement in
practice than just comparing physical addresses, this is needed for not leaking
information about pointers (see \Cref{sec:observing-pointers}).
The special expression $\coffset$ extracts the offset component of a pointer;
we introduce it to illustrate that for satisfying our memory characterization
pointer offsets do not need to be hidden (as opposed to block identifiers).
The less-than-or-equal operator only applies to numbers---in particular,
pointers cannot be compared. However, since we can extract pointer offsets, we
can compare those instead.
\iflater
\bcp{Why do we need $\coffset$?}\ch{Explained above the reason I think we have
  $\coffset$; it's a quite artificial reason, and the main disadvantage is that
  it's not really implementable in most real systems.}\ch{Another way to
  illustrate could be to allow pointers to the same region to be compared (both
  via less than and via subtraction).}\ch{Even for micro-policies, comparing
  pointers to the same region can be done with a normal instruction, while
  get offset would need to be a monitor call.}
%
\fi

The definition of command evaluation employs an auxiliary partial function
that computes the result of evaluating a program along with the set of block
identifiers that were allocated during evaluation.  Formally,
$\lsb c\rsb_{+} : \St \partfun \Ot_{+}$, where $\Ot_{+}$ is an extended
set of outcomes defined as $\power_\fin(\I) \times \St \uplus \{\oerror\}$.
We then set
\begin{align*}
  \lsb c \rsb(l, m) & =
  \begin{cases}
    (l', m') & \text{if $\lsb c\rsb_{+}(l, m) = (I, l', m')$} \\
    \oerror  & \text{if $\lsb c\rsb_{+}(l, m) = \oerror$} \\
    \bot  & \text{if $\lsb c\rsb_{+}(l, m) = \bot$}
  \end{cases} \\
  \finalids(l, m) & =
  \begin{cases}
    \ids(l, m) \setminus I & \text{if $\lsb c\rsb_{+}(l, m) = (I, l', m')$} \\
    \emptyset  & \text{otherwise}
  \end{cases}
\end{align*}

To define $\lsb c \rsb_{+}$, we first endow the set
$\St \partfun \Ot_{+}$ with the partial order of program approximation:
\[ f \sqsubseteq g \ \ \teq\ \ \forall s, f(s) \neq \bot \Rightarrow f(x) = g(x) \]
This allows us to define the semantics of iteration (the rule for
$\cwhiledo{e}{c}$) in a standard way using the Kleene fixed point operator
$\codeface{fix}$.

The definition of $\lsb c \rsb_{+}$ appears in \Cref{fig:command-sem}, where
several of the rules use a $\bind$ operator (\Cref{fig:bind}) to
manage the ``plumbing'' of the sets of allocated block ids between the
evaluation of one subcommand and the next.  The rules for $\cif$ and $\cwhile$
also use an auxiliary operator $\cif$ (also defined in \Cref{fig:bind}) that
turns non-boolean guards into errors.

The evaluation rules for $\cskip$, sequencing, conditionals, $\codeface{while}$,
and assignment are standard.  The rule for heap lookup, $x \gets [e]$, evaluates
$e$ to a pointer and then looks it up in the heap, yielding an error if $e$ does
not evaluate to a pointer or if it evaluates to a pointer that is invalid,
either because its block id is not allocated or because its offset is out of
bounds.  Similarly, the heap mutation command, $[e_1] \gets e_2$, requires that
$e_1$ evaluate to a pointer that is valid in the current memory $m$ (i.e., such
that looking it up in $m$ yields something other than $\bot$).  The allocation
command $x \gets \calloc(e)$ first evaluates $e$ to an integer $n$, then calculates
the next free block id for the current machine state ($\freshf(\ids(l, m))$); it
yields a new machine state where $x$ points to the first cell in the new block
and where a new block of $n$ cells is added the heap, all initialized to $0$.
Finally, $\cfree(e)$ evaluates $e$ to a pointer and yields a new heap where
every cell sharing the same block id as this pointer is undefined.

\clearpage
\bibliographystyle{plainurl.bst}
\bibliography{refs,mp,safe}

\begin{thebibliography}{\tiny10}

\bibitem{caja}
\href{http://code.google.com/p/google-caja/wiki/AttackVectors}{Caja. {Attack}
  vectors for privilege escalation}, 2012.

\bibitem{Agten0P15}
P.~Agten, B.~Jacobs, and F.~Piessens.
\newblock
  \href{https://lirias.kuleuven.be/bitstream/123456789/471365/3/sound-verification.pdf}{Sound
  modular verification of {C} code executing in an unverified context}.
\newblock \iffull{In {\em 42nd Annual {ACM} {SIGPLAN-SIGACT} Symposium on
  Principles of Programming Languages}}\else{{\em POPL}}\fi{}. 2015.

\bibitem{AhmedFM07}
A.~Ahmed, M.~Fluet, and G.~Morrisett.
\newblock
  \href{http://content.iospress.com/articles/fundamenta-informaticae/fi77-4-06}{L\({}^{\mbox{3}}\):
  {A} linear language with locations}.
\newblock {\em Fundam. Inform.}, 77(4):397--449, 2007.

\bibitem{askarov08:tini_leaks_more_than_1_bit}
A.~Askarov, S.~Hunt, A.~Sabelfeld, and D.~Sands.
\newblock
  \href{http://www.cse.chalmers.se/~andrei/esorics08.pdf}{Termination-insensitive
  noninterference leaks more than just a bit}.
\newblock \iffull{In {\em 13th European Symposium on Research in Computer
  Security (ESORICS)}}\else{{\em ESORICS}}\fi{}. 2008.

\bibitem{micropolicies2015}
A.~{Azevedo de Amorim}, M.~D\'en\`es, N.~Giannarakis, C.~Hri\c{t}cu, B.~C.
  Pierce, A.~{Spector-Zabusky}, and A.~Tolmach.
\newblock
  \href{http://prosecco.gforge.inria.fr/personal/hritcu/publications/micro-policies.pdf}{Micro-policies:
  Formally verified, tag-based security monitors}.
\newblock \iffull{In {\em 36th IEEE Symposium on Security and Privacy (Oakland
  S\&P)}}\else{{\em Oakland S\&P}}\fi{}. 2015.

\bibitem{BackesKR09}
M.~Backes, B.~K{\"{o}}pf, and A.~Rybalchenko.
\newblock \href{https://doi.org/10.1109/SP.2009.18}{Automatic discovery and
  quantification of information leaks}.
\newblock \iffull{In {\em 30th {IEEE} Symposium on Security and Privacy (S{\&}P
  2009), 17-20 May 2009, Oakland, California, {USA}}}\else{{\em Oakland
  S\&P}}\fi{}, 2009.

\bibitem{BalabonskiPP14}
T.~Balabonski, F.~Pottier, and J.~Protzenko.
\newblock \href{http://dx.doi.org/10.1007/978-3-319-07151-0\_16}{Type soundness
  and race freedom for {M}ezzo}.
\newblock \iffull{In M.~Codish and E.~Sumii, editors, {\em Functional and Logic
  Programming -- 12th International Symposium, {FLOPS} 2014, Kanazawa, Japan,
  June 4-6, 2014. Proceedings}}\else{{\em FLOPS}}\fi{}. 2014.

\bibitem{Benton:2009}
N.~Benton and N.~Tabareau.
\newblock
  \href{http://dblp.uni-trier.de/db/conf/tldi/tldi2009.html#BentonT09}{Compiling
  functional types to relational specifications for low level imperative code.}
\newblock In A.~Kennedy and A.~Ahmed, editors, {\em TLDI}. 2009.

\bibitem{BhargavanDM13}
K.~Bhargavan, A.~Delignat{-}Lavaud, and S.~Maffeis.
\newblock \href{http://dx.doi.org/10.1007/978-3-319-10082-1_4}{Defensive
  {JavaScript} - building and verifying secure web components}.
\newblock \iffull{In A.~Aldini, J.~Lopez, and F.~Martinelli, editors, {\em
  Foundations of Security Analysis and Design {VII} - {FOSAD} 2012/2013
  Tutorial Lectures}}\else{{\em FOSAD}}\fi{}. 2013.

\bibitem{cheri_asplos2015}
D.~Chisnall, C.~Rothwell, R.~N.~M. Watson, J.~Woodruff, M.~Vadera, S.~W. Moore,
  M.~Roe, B.~Davis, and P.~G. Neumann.
\newblock
  \href{https://www.cl.cam.ac.uk/~dc552/papers/asplos15-memory-safe-c.pdf}{Beyond
  the {PDP-11}: Architectural support for a memory-safe {C} abstract machine}.
\newblock \iffull{In {\em Proceedings of the Twentieth International Conference
  on Architectural Support for Programming Languages and Operating
  Systems}}\else{{\em ASPLOS}}\fi{}. 2015.

\bibitem{ClauseDOP07}
J.~A. Clause, I.~Doudalis, A.~Orso, and M.~Prvulovic.
\newblock
  \href{http://www.cc.gatech.edu/~orso/papers/clause.doudalis.orso.prvulovic.pdf}{Effective
  memory protection using dynamic tainting}.
\newblock \iffull{In {\em 22nd IEEE/ACM International Conference on Automated
  Software Engineering (ASE)}}\else{{\em ASE}}\fi{}. 2007.

\bibitem{AmorimCDDHPPPT16}
A.~A. de~Amorim, N.~Collins, A.~DeHon, D.~Demange, C.~Hritcu, D.~Pichardie,
  B.~C. Pierce, R.~Pollack, and A.~Tolmach.
\newblock \href{http://dx.doi.org/10.3233/JCS-15784}{A verified
  information-flow architecture}.
\newblock {\em Journal of Computer Security}, 24(6):689--734, 2016.

\bibitem{DeviettiBMZ08}
J.~Devietti, C.~Blundell, M.~M.~K. Martin, and S.~Zdancewic.
\newblock
  \href{http://acg.cis.upenn.edu/papers/asplos08_hardbound.pdf}{{HardBound}:
  Architectural support for spatial safety of the {C} programming language}.
\newblock \iffull{In {\em 13th International Conference on Architectural
  Support for Programming Languages and Operating Systems}}\else{{\em
  ASPLOS}}\fi{}, 2008.

\bibitem{DevriesePB16}
D.~Devriese, F.~Piessens, and L.~Birkedal.
\newblock
  \href{http://cs.au.dk/~birke/papers/object-capabilities-tr.pdf}{Reasoning
  about object capabilities with logical relations and effect parametricity}.
\newblock \iffull{In {\em 1st IEEE European Symposium on Security and
  Privacy}}\else{{\em EuroS\&P}}\fi{}, 2016.

\bibitem{pump_asplos2015}
U.~Dhawan, C.~Hri\c{t}cu, R.~Rubin, N.~Vasilakis, S.~Chiricescu, J.~M. Smith,
  T.~F. {Knight, Jr.}, B.~C. Pierce, and A.~DeHon.
\newblock
  \href{http://ic.ese.upenn.edu/abstracts/sdmp_asplos2015.html}{Architectural
  support for software-defined metadata processing}.
\newblock \iffull{In {\em International Conference on Architectural Support for
  Programming Languages and Operating Systems}}\else{{\em ASPLOS}}\fi{}, 2015.

\bibitem{DurumericKAHBLWABPP14}
Z.~Durumeric, J.~Kasten, D.~Adrian, J.~A. Halderman, M.~Bailey, F.~Li,
  N.~Weaver, J.~Amann, J.~Beekman, M.~Payer, and V.~Paxson.
\newblock \href{http://doi.acm.org/10.1145/2663716.2663755}{The matter of
  {Heartbleed}}.
\newblock \iffull{In C.~Williamson, A.~Akella, and N.~Taft, editors, {\em
  Proceedings of the 2014 Internet Measurement Conference, {IMC} 2014,
  Vancouver, BC, Canada, November 5-7, 2014}}\else{{\em IMC}}\fi{}. 2014.

\bibitem{ElliottPWHBSSL15}
T.~Elliott, L.~Pike, S.~Winwood, P.~C. Hickey, J.~Bielman, J.~Sharp, E.~L.
  Seidel, and J.~Launchbury.
\newblock \href{https://www.cs.indiana.edu/~lepike/pubs/ivory.pdf}{Guilt free
  {Ivory}}.
\newblock \iffull{In B.~Lippmeier, editor, {\em Proceedings of the 8th {ACM}
  {SIGPLAN} Symposium on Haskell, Haskell 2015, Vancouver, BC, Canada,
  September 3-4, 2015}}\else{{\em Haskell}}\fi{}. 2015.

\bibitem{FournetSCDSL13}
C.~Fournet, N.~Swamy, J.~Chen, P.-{\'E}. Dagand, P.-Y. Strub, and B.~Livshits.
\newblock \href{https://research.microsoft.com/pubs/176601/js-star.pdf}{Fully
  abstract compilation to {JavaScript}}.
\newblock \iffull{In {\em 40th Annual {ACM} {SIGPLAN-SIGACT} Symposium on
  Principles of Programming Languages}}\else{{\em POPL}}\fi{}. 2013.

\bibitem{GoguenM82}
J.~A. Goguen and J.~Meseguer.
\newblock
  \href{http://spy.sci.univr.it/papers/Isa-orig/Sicurezza/NonInterferenza/noninter.pdf}{Security
  policies and security models}.
\newblock \iffull{In {\em Symposium on Security and Privacy}}\else{{\em
  S\&P}}\fi{}, 1982.

\bibitem{Hicks:memory-safety}
M.~Hicks.
\newblock What is memory safety?
\newblock \url{http://www.pl-enthusiast.net/2014/07/21/memory-safety/}, 2014.

\bibitem{ISO:C99}
ISO.
\newblock
  \href{http://www.open-std.org/jtc1/sc22/wg14/www/docs/n1124.pdf}{{ISO} {C}
  standard 1999}.
\newblock Technical report, ISO, 1999.
\newblock ISO/IEC 9899:1999 draft.

\bibitem{JanaS12a}
S.~Jana and V.~Shmatikov.
\newblock \href{https://doi.org/10.1109/SP.2012.19}{Memento: Learning secrets
  from process footprints}.
\newblock \iffull{In {\em {IEEE} Symposium on Security and Privacy, {SP} 2012,
  21-23 May 2012, San Francisco, California, {USA}}}\else{{\em Oakland
  S{\&}P}}\fi{}. 2012.

\bibitem{KangHMGZV15}
J.~Kang, C.~Hur, W.~Mansky, D.~Garbuzov, S.~Zdancewic, and V.~Vafeiadis.
\newblock \href{https://www.seas.upenn.edu/~wmansky/mcast.pdf}{A formal {C}
  memory model supporting integer-pointer casts}.
\newblock \iffull{In D.~Grove and S.~Blackburn, editors, {\em Proceedings of
  the 36th {ACM} {SIGPLAN} Conference on Programming Language Design and
  Implementation, Portland, OR, USA, June 15-17, 2015}}\else{{\em PLDI}}\fi{}.
  2015.

\bibitem{Krebbers15}
R.~Krebbers.
\newblock \href{http://robbertkrebbers.nl/research/thesis.pdf}{{\em The {C}
  standard formalized in {Coq}}}.
\newblock PhD thesis, Radboud University Nijmegen, 2015.

\bibitem{LowFat2013}
A.~Kwon, U.~Dhawan, J.~M. Smith, T.~F. {Knight, Jr.}, and A.~DeHon.
\newblock \href{http://www.crash-safe.org/node/27}{Low-fat pointers: compact
  encoding and efficient gate-level implementation of fat pointers for spatial
  safety and capability-based security}.
\newblock \iffull{In {\em ACM SIGSAC Conference on Computer and Communications
  Security (CCS)}}\else{{\em CCS}}\fi{}. 2013.

\bibitem{LeroyB08}
X.~Leroy and S.~Blazy.
\newblock
  \href{http://pauillac.inria.fr/~xleroy/publi/memory-model-journal.pdf}{Formal
  verification of a {C}-like memory model and its uses for verifying program
  transformations}.
\newblock \iffull{{\em Journal of Automated Reasoning}}\else{{\em JAR}}\fi{},
  41(1):1--31, 2008.

\bibitem{MaffeisMT10}
S.~Maffeis, J.~C. Mitchell, and A.~Taly.
\newblock \href{https://www.doc.ic.ac.uk/~maffeis/papers/oakland10.pdf}{Object
  capabilities and isolation of untrusted web applications}.
\newblock \iffull{In {\em 31st {IEEE} Symposium on Security and Privacy, S{\&}P
  2010, 16-19 May 2010, Berleley/Oakland, California, {USA}}}\else{{\em Oakland
  S{\&}P}}\fi{}, 2010.

\bibitem{MemarianMLNCWS16}
K.~Memarian, J.~Matthiesen, J.~Lingard, K.~Nienhuis, D.~Chisnall, R.~N.~M.
  Watson, and P.~Sewell.
\newblock \href{http://doi.acm.org/10.1145/2908080.2908081}{Into the depths of
  {C}: elaborating the de facto standards}.
\newblock \iffull{In {\em Proceedings of the 37th {ACM} {SIGPLAN} Conference on
  Programming Language Design and Implementation, {PLDI} 2016, Santa Barbara,
  CA, USA, June 13-17, 2016}}\else{{\em PLDI}}\fi{}, 2016.

\bibitem{MeyerovichL10}
L.~A. Meyerovich and V.~B. Livshits.
\newblock \href{http://dx.doi.org/10.1109/SP.2010.36}{Conscript: Specifying and
  enforcing fine-grained security policies for javascript in the browser}.
\newblock \iffull{In {\em 31st {IEEE} Symposium on Security and Privacy, S{\&}P
  2010, 16-19 May 2010, Berleley/Oakland, California, {USA}}}\else{{\em Oakland
  S{\&}P}}\fi{}, 2010.

\bibitem{Morrisett:1995}
G.~Morrisett, M.~Felleisen, and R.~Harper.
\newblock \href{http://doi.acm.org/10.1145/224164.224182}{Abstract models of
  memory management}.
\newblock \iffull{In {\em Proceedings of the Seventh International Conference
  on Functional Programming Languages and Computer Architecture}}\else{{\em
  FPCA}}\fi{}. 1995.

\bibitem{NagarakatteZMZ09}
S.~Nagarakatte, J.~Zhao, M.~M.~K. Martin, and S.~Zdancewic.
\newblock
  \href{http://repository.upenn.edu/cgi/viewcontent.cgi?article=1941&context=cis_reports}{{SoftBound}:
  highly compatible and complete spatial memory safety for {C}}.
\newblock \iffull{In {\em ACM SIGPLAN Conference on Programming Language Design
  and Implementation (PLDI)}}\else{{\em PLDI}}\fi{}. 2009.

\bibitem{NagarakatteZMZ10}
S.~Nagarakatte, J.~Zhao, M.~M.~K. Martin, and S.~Zdancewic.
\newblock \href{http://acg.cis.upenn.edu/papers/ismm10_cets.pdf}{{CETS}:
  compiler enforced temporal safety for {C}}.
\newblock \iffull{In {\em 9th International Symposium on Memory
  Management}}\else{{\em ISMM}}\fi{}. 2010.

\bibitem{ccured_toplas2005}
G.~C. Necula, J.~Condit, M.~Harren, S.~McPeak, and W.~Weimer.
\newblock \href{http://doi.acm.org/10.1145/1065887.1065892}{{CCured}: type-safe
  retrofitting of legacy software}.
\newblock \iffull{{\em ACM Trans. Program. Lang. Syst.}}\else{{\em
  TOPLAS}}\fi{}, 27(3):477--526, 2005.

\bibitem{Pitts:2013}
A.~M. Pitts.
\newblock {\em Nominal Sets: Names and Symmetry in Computer Science}.
\newblock Cambridge University Press, New York, NY, USA, 2013.

\bibitem{pottier-protzenko-13}
F.~Pottier and J.~Protzenko.
\newblock Programming with permissions in {Mezzo}.
\newblock \iffull{In {\em Proceedings of the 2013 {ACM} {SIGPLAN} International
  Conference on Functional Programming (ICFP'13)}}\else{{\em ICFP}}\fi{}, 2013.

\bibitem{Reynolds:2002}
J.~C. Reynolds.
\newblock \href{http://dl.acm.org/citation.cfm?id=645683.664578}{Separation
  logic: A logic for shared mutable data structures}.
\newblock \iffull{In {\em Proceedings of the 17th Annual IEEE Symposium on
  Logic in Computer Science}}\else{{\em LICS}}\fi{}. 2002.

\bibitem{Rust}
The {Rust} programming language.
\newblock \url{http://www.rust-lang.org}, 2017.

\bibitem{SchlesingerPSWZ14}
C.~Schlesinger, K.~Pattabiraman, N.~Swamy, D.~Walker, and B.~G. Zorn.
\newblock \href{http://dx.doi.org/10.3233/JCS-140502}{Modular protections
  against non-control data attacks}.
\newblock \iffull{{\em Journal of Computer Security}}\else{{\em JCS}}\fi{},
  22(5):699--742, 2014.

\bibitem{Smith09}
G.~Smith.
\newblock \href{https://doi.org/10.1007/978-3-642-00596-1_21}{On the
  foundations of quantitative information flow}.
\newblock \iffull{In L.~de~Alfaro, editor, {\em Foundations of Software Science
  and Computational Structures, 12th International Conference, {FOSSACS} 2009,
  Held as Part of the Joint European Conferences on Theory and Practice of
  Software, {ETAPS} 2009, York, UK, March 22-29, 2009. Proceedings}}\else{{\em
  FoSSaCS}}\fi{}. 2009.

\bibitem{StefanBYLTRM13}
D.~Stefan, P.~Buiras, E.~Z. Yang, A.~Levy, D.~Terei, A.~Russo, and
  D.~Mazi{\`e}res.
\newblock
  \href{http://www.scs.stanford.edu/~deian/pubs//stefan:2013:eliminating.pdf}{Eliminating
  cache-based timing attacks with instruction-based scheduling}.
\newblock \iffull{In {\em 18th European Symposium on Research in Computer
  Security (ESORICS)}}\else{{\em ESORICS}}\fi{}. 2013.

\bibitem{SwamyHMGJ06}
N.~Swamy, M.~W. Hicks, G.~Morrisett, D.~Grossman, and T.~Jim.
\newblock \href{http://www.cs.umd.edu/~mwh/papers/cyc-mm-scp.pdf}{Safe manual
  memory management in {Cyclone}}.
\newblock \iffull{{\em Science of Computer Programming}}\else{{\em SCP}}\fi{},
  62(2):122--144, 2006.

\bibitem{SwaseyGD17}
D.~Swasey, D.~Garg, and D.~Dreyer.
\newblock \href{https://people.mpi-sws.org/~swasey/papers/ocpl}{Robust and
  compositional verification of object capability patterns}.
\newblock To appear at OOPSLA, 2017.

\bibitem{Szekeres2013}
L.~Szekeres, M.~Payer, T.~Wei, and D.~Song.
\newblock \href{http://lenx.100871.net/papers/War-oakland-CR.pdf}{{SoK}:
  Eternal war in memory}.
\newblock \iffull{In {\em IEEE Symposium on Security and Privacy}}\else{{\em
  IEEE S\&P}}\fi{}. 2013.

\bibitem{TalyEMMN11}
A.~Taly, {\'{U}}.~Erlingsson, J.~C. Mitchell, M.~S. Miller, and J.~Nagra.
\newblock \href{http://dx.doi.org/10.1109/SP.2011.39}{Automated analysis of
  security-critical javascript apis}.
\newblock \iffull{In {\em 32nd {IEEE} Symposium on Security and Privacy, S{\&}P
  2011, 22-25 May 2011, Berkeley, California, {USA}}}\else{{\em Oakland
  S{\&}P}}\fi{}. 2011.

\bibitem{Turon17}
A.~Turon.
\newblock \href{http://dl.acm.org/citation.cfm?id=3011999}{Rust: from {POPL} to
  practice (keynote)}.
\newblock \iffull{In G.~Castagna and A.~D. Gordon, editors, {\em Proceedings of
  the 44th {ACM} {SIGPLAN} Symposium on Principles of Programming Languages,
  {POPL} 2017, Paris, France, January 18-20, 2017}}\else{{\em POPL}}\fi{}.
  2017.

\bibitem{m7negative}
C.~Williams.
\newblock
  \href{http://www.theregister.co.uk/2015/10/28/oracle_sparc_m7/}{Oracle's
  {Larry Ellison} claims his {Sparc M7} chip is hacker-proof -- errr...}
\newblock The Register, 2015.

\bibitem{Yang:2014}
E.~Z. Yang and D.~Mazi\`{e}res.
\newblock \href{http://doi.acm.org/10.1145/2594291.2594341}{Dynamic space
  limits for {Haskell}}.
\newblock \iffull{In {\em Proceedings of the 35th ACM SIGPLAN Conference on
  Programming Language Design and Implementation}}\else{{\em PLDI}}\fi{}. 2014.

\bibitem{Yang:2002}
H.~Yang and P.~W. O'Hearn.
\newblock \href{http://dl.acm.org/citation.cfm?id=646794.704850}{A semantic
  basis for local reasoning}.
\newblock \iffull{In {\em Proceedings of the 5th International Conference on
  Foundations of Software Science and Computation Structures}}\else{{\em
  FoSSaCS}}\fi{}. 2002.

\bibitem{ZhangAM12}
D.~Zhang, A.~Askarov, and A.~C. Myers.
\newblock \href{http://doi.acm.org/10.1145/2254064.2254078}{Language-based
  control and mitigation of timing channels}.
\newblock \iffull{In J.~Vitek, H.~Lin, and F.~Tip, editors, {\em {ACM}
  {SIGPLAN} Conference on Programming Language Design and Implementation,
  {PLDI} '12, Beijing, China - June 11 - 16, 2012}}\else{{\em PLDI}}\fi{}.
  2012.

\end{thebibliography}

\end{document}